\documentclass[twocolumn,showpacs,preprintnumbers,amsmath,amssymb,superscriptaddress, floatfix]{revtex4-2}

\usepackage[maxfloats=256]{morefloats}

\usepackage{graphicx}% Include figure files
\usepackage{dcolumn}% Align table columns on the decimal point
\usepackage{bm}% bold math
\usepackage{tabularx}
\usepackage{ulem}
\newcolumntype{M}{>{\centering\arraybackslash}m{1.85cm}}
\usepackage[export]{adjustbox}
\usepackage{float}
\usepackage{amsmath}
\usepackage{color}   %May be necessary if you want to color links
\usepackage{hyperref}
\usepackage{comment}
\hypersetup{
	colorlinks=true, %set true if you want colored links
	linktoc=all,     %set to all if you want both sections and subsections linked
	linkcolor=blue,  % Choose some color if you want links to stand out
}

\makeatletter
\newcommand{\colorcaption}[2][]{%
	\begingroup%
	\renewcommand{\@caption@fignum@sep}{ (Color online). }%
	\caption[#1]{#2}%
	\endgroup%
}
\makeatother

\newcommand\T{\rule{0pt}{3ex}}       % Top strut
\newcommand\B{\rule[-1.5ex]{0pt}{0pt}} % Bottom strut

\newcommand{\orcid}[1]{\href{https://orcid.org/#1}{\hskip2pt\includegraphics[width=9pt]{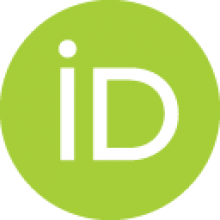}}}

\begin{document}

 %	\title{ Electron spectral shapes for forbidden non-unique $\beta^{-}$  transitions with  A=85-123}
    	\title{ Forbidden non-unique $\beta^{-}$  transitions and $g_A$-sensitive $\beta$ spectral shapes}

	\author{Archana Saxena\orcid{0000-0001-5175-6041}}
 \email{asaxena@ph.iitr.ac.in}
	\address{Department of Physics, Indian Institute of Technology Roorkee, Roorkee 247667, India}
     \author{Praveen C. Srivastava\orcid{0000-0001-8719-1548}}
\email{praveen.srivastava@ph.iitr.ac.in}
	\address{Department of Physics, Indian Institute of Technology Roorkee, Roorkee 247667, India}

	\date{\hfill \today}

	\begin{abstract}
In the present work, we have done a systematic study of beta decay properties such as  electron spectral-shapes, shape factors, and log$ft$ values for the higher forbidden non-unique $\beta^{-}$ transitions in the mass region A=85-123. We have  performed the nuclear shell model (SM) calculations to explore the sensitivity of the electron spectral-shapes for different axial-vector coupling constants $g_{A}=0.8-1.27$. The effective interactions GWBXG, G-matrix, SNET and SN100PN are used for different model spaces. In the present work, we have computed the electron spectral-shapes of $^{85}$Br, $^{87}$Rb, $^{93}$Zr, $^{97}$Zr, $^{101}$Mo, $^{115}$Cd, $^{117}$Cd, $^{119}$In, $^{123}$Sn and $^{135}$Cs by constraining the small relativistic nuclear matrix element from conserved vector-current hypothesis (CVC). We have found that the electron spectral-shapes are strongly dependent on  $g_{A}$  except the second forbidden non-unique $\beta^{-}$ transition $^{93}$Zr.

 \end{abstract}
\maketitle

\section{Introduction}

In recent years, interest in studying electron spectral-shapes has grown significantly from both experimental and theoretical perspectives. The electron spectral-shapes  which are sensitive to axial coupling constants ($g_{A}$) play crucial role in rare-events experiments. Also, the theoretical prediction of $g_{A}$ becomes important while designing the experiment to detect the neutrinoless double beta decay ($0v\beta\beta$) events and  in predicting $\beta$-decay rates. The single $\beta$-decay rates are proportional to the second power of $g_{A}$ whereas for double  $\beta$-decay are proportional to the fourth power of $g_{A}$, 
thus the precise value of $g_{A}$ should be taken \cite{EJIRI20191, Suhonen1}.
%but the what value of $g_{A}$ should be taken is not clear. 
 A comprehensive study was previously conducted on allowed and forbidden $\beta$-decays within the framework of the nuclear shell model (SM) and microscopic quasiparticle-phonon model (MQPM) to get the information related to the effective value of $g_{A}$ \cite{Suhonen1, Joel_2017, Joel_2017_2}.

From the standard model, the conserved vector current (CVC) and partially conserved axial-vector-current (PCAC) hypotheses yield free nucleon value of vector coupling constant $g_{V} = 1.00$ and axial-vector coupling constant $g_{A} = 1.27$, respectively \cite{ED2007}. However, inside nuclear matter, many-body correlations affect the value of $g_{A}$, and a quenched value may be required to reproduce experimental data \cite{jouni}. {\color{black} In  Ref. \cite{Gysbers}, ab initio calculations have been done for selected lighter nuclei and two-body current is included. However, it is very difficult to do ab initio calculations for heavier nuclei. So, SM calculations in a truncated model space are desirable, which requires quenching in $g_{A}$. 
The SSM has recently gained popularity for investigating the effective value of $g_{A}$.} 
In current scenario, there are two methods to extract the effective of $g_{A}$: (i) the half-life  method \cite{Martinez1996, Akumar1,Vkumar1}, and (ii) the spectrum shape method (SSM). The effective value of $g_{A}$ has previously been investigated using a half-life  method  for the different values of $g_{A}$, confirming the quenching of $g_{A}$ \cite{Ejiri2014, Ejiri2015}. 
In the SSM method, the shapes of computed spectra at different $g_{A}$ values are compared with the experimental electron spectra to extract the effective value of $g_{A}$ \cite{mika2016}.
Such method is used recently  for the fourth forbidden non-unique $\beta$-decay of $^{113}$Cd and $^{115}$In \cite{mika2016,Bodenstein2020, joel2021, Leder2022} and measurements of $^{113}$Cd and $^{115}$In spectra are being also extended to other potentially sensitive candidates \cite{Pagnanini 2023}. For $^{113}$Cd, with the SSM, the  MQPM, SM, and the microscopic interacting boson-fermion model (IBFM-2) \cite{Iachello} yield the ratio of axial-vector coupling constant to vector coupling constant approximately $0.92$ when compared with the experimental spectra.  This value was the closest agreement among all three models which signifies that the SSM method does not strongly depend on the details of the Hamiltonian or many-body methods \cite{Kotila, Belli2007, mika2016} while using half-life method the $g_{A}$ value is not consistent with all three models.
Recently,  SSM method is enhanced after constraining the  relativistic nuclear matrix element (s-NME) from CVC theory and obtained the better picture of electron spectral-shapes of $sd$ shell nuclei $^{36}$Cl, and $^{24}$Na \cite{Anil2020PRC}. 
Along with this, the spectral shapes of the second forbidden non-unique beta decay of $^{59,60}$Fe are also studied in the $fp$ shell with different  $g_{A}$  and the strong $g_{A}$ dependency was found which shows $^{59,60}$Fe are the best candidates for SSM and useful for shape measurement experiments \cite{anil2021}.

Recently, the realistic shell model (RSM) calculations have been performed using bare and effective operators to predict the log$ft$ values and energy spectra of the second forbidden $\beta^{-}$ decay of $^{94}$Nb and $^{99}$Tc and the fourth forbidden $\beta^{-}$ decay of $^{113}$Cd and $^{115}$In \cite{Gregorio2024}. Previously, the renormalization of Hamiltonian and transition operator within the RSM is also studied for two neutrino mode of double beta decay (2$\nu\beta\beta$) and  0$\nu\beta\beta$ \cite{Coraggio2019, Coraggio2022}.
Along with this, a recent nuclear SM study on $g_{A}$ sensitive $\beta^{-}$ spectral shapes in the mass region A=86-99 is done where the $\beta^{-}$ transitions are potentially of great interest for future rare events experiments \cite{Ramalho2024} which demand the beyond standard model (BSM) physics by measurements of rare nuclear $\beta$ decays and double $\beta$ decays. Recently, for the heavier nuclei  $^{212,214}$Pb and $^{212,214}$Bi, a systematic study of electron spectra has been performed within the framework of shell model Hamiltonian which plays an important role in neutrino and dark-matter experiments due to their background contribution in these experiments. In this study there was no strong dependency of $g_{A}$ on spectral shape was found but a moderate dependency was obtained on small relativistic nuclear matrix elements (s-NMEs) \cite{Ramalho}. Although, s-NME is small, it can have a substantial effect on the partial half-life and the shape of beta electron spectra \cite{ joel2021, joel2023}.
%In this continuation, using shell model
In the present work using shell model, 
we have explored the sensitivity of electron spectral-shapes of $^{85}$Br, $^{87}$Rb, $^{93}$Zr, $^{97}$Zr, $^{101}$Mo, $^{115}$Cd, $^{117}$Cd, $^{119}$In, $^{123}$Sn and $^{135}$Cs by constraining the small relativistic nuclear matrix element from conserved vector-current hypothesis (CVC). The results are reported corresponding to different $g_{A}$ values.\\

The present manuscript is organized as follows: In section \ref{formalism}, we give the theoretical background  behind the SM and beta spectrum. In section \ref{result}, we present the result and discussion for  non-unique $\beta^{-}$ transitions. In the end, we conclude the present work in section \ref{Conclusion}.

\section{Theoretical Framework} \label{formalism}

\subsection{Formalism} \label{beta}

In the present study, we have computed the beta spectrum for forbidden non-unique $\beta^{-}$ decay transitions, analyzing the energy distribution of beta electrons emitted in $\beta^{-}$ decays. Based on the impulse approximation in nuclear $\beta$-decay, the generalized theoretical formalism (Behrens-B$\ddot{\rm u}$hring) can be found in Refs. \cite{Kotila, mst2006, behrens1982, hfs1966}, which is used in the present work. Apart from this formalism, other  $\beta$-decay formalisms such as Holstein's form factors \cite{Holstein1974} and Donnelly and Walecka formalism have been also introduced \cite{Connell1972, DONNELLY1972275, DONNELLY197381, DONNELLY1976368, DONNELLY19791,Walecka}. The Donnelly and Walecka formalism is widely used in the BSM and suitable in the modern $ab~initio$ calculations and this formalism is quite useful for current searches of BSM signatures in precision $\beta^{-}$ decay experiments \cite{Ayala2022}. In the present work, we have followed the Behrens-B$\ddot{\rm u}$hring formalism.
The  partial half-life  of  $\beta$-decay is given as \cite{Kotila, behrens1982}

    \begin{equation}
	t_{1/2}= \frac{\kappa}{\tilde{C}}.
        \end{equation}
 
	Here, $\tilde{C}$ represents the dimensionless integrated shape function, while $\kappa$, which includes natural constants, has a value of 6289 s \cite{Patrignani}. Furthermore, $\tilde{C}$ can be calculated as follows
  
  \begin{eqnarray}\label{eq2}
\label{tc}
\tilde{C}=\int _{1}^{w_{0}}C(w_{e})pw_{e}(w_{0}-w_{e})^{2}F_{0}(Z,w_{e})dw_{e}.
\end{eqnarray}
In Eq. \ref{eq2}, we use the usual dimensionless kinematics quantities which are divided by the electron rest mass as $w_0=W_0/m_ec^2$, $w_e=W_e/m_ec^2$, and $p=p_ec/m_ec^2=\sqrt{(w_e^2-1)}$.
The  $w_e$  and $p_e$ are the energy and momentum of emitted electrons and the $w_0$ is the endpoint energy, respectively. The $F_0(Z, w_e)$ is the Fermi function where $Z$ is the atomic number of the daughter nucleus. 

The shape factor $C(w_e)$ in Eq. \ref{eq2} is given as  

 \begin{eqnarray} \label{eq3}
%\begin{split}
C(w_e)  = \sum_{k_e,k_\nu,K}\lambda_{k_e} \Big[M_K(k_e,k_\nu)^2+m_K(k_e,k_\nu)^2 \nonumber\\
    -\frac{2\gamma_{k_e}}{k_ew_e}M_K(k_e,k_\nu)m_K(k_e,k_\nu)\Big].
%\end{split}
\end{eqnarray}
 Here, $K$ is the order of forbiddenness of the $\beta$ decay, $k_e$ and $k_\nu$ (taking values 1, 2, 3,...) are positive integers related to the partial-wave expansion of the leptonic wave functions.
The expressions for $M_K(k_e,k_\nu)$ and $m_K(k_e,k_\nu)$ involve different nuclear matrix elements (NMEs) and kinematic factors \cite{Kotila, behrens1982}. 
Here, $\gamma_{k_e} = \sqrt{k_e^2 - (\alpha Z)^2}$ and $y = (\alpha Z w_e / p_e c)$ are auxiliary quantities, where $\alpha = 1/137$ is the fine-structure constant. The term $\lambda_{k_e} = {F_{k_e-1}(Z, w_e)}/{F_0(Z, w_e)}$ represents the Coulomb function, where $F_{k_e-1}(Z, w_e)$ is the generalized Fermi function \cite{Kotila, mst2006, Anil2020PRC}, which is given as below

\begin{eqnarray} \label{eq4}
F_{k_e-1}(Z,w_e) &=4^{k_e-1}(2k_e)(k_e+\gamma_{k_e})[(2k_e-1)!!]^2e^{\pi{y}} \nonumber \\
 & \times\left(\frac{2p_eR}{\hbar}\right)^{2(\gamma_{k_e}-k_e)}\left(\frac{|\Gamma(\gamma_{k_e}+iy)|}{\Gamma(1+2\gamma_{k_e})}\right)^2.
\end{eqnarray}
 
The NMEs, which contain nuclear structure information for a transition between an initial (i) and final (f) state, can be written as:

\begin{align}
\begin{split}
^{\rm V/A}\mathcal{M}_{KLS}^{(N)}(pn)(k_e,m,n,\rho)& \\ =\frac{\sqrt{4\pi}}{\widehat{J}_i}
\sum_{pn} \, ^{\rm V/A}m_{KLS}^{(N)}(pn)(&k_e,m,n,\rho)(\Psi_f|| [c_p^{\dagger}
\tilde{c}_n]_K || \Psi_i).
\label{eq5}
\end{split}
\end{align}
The term $(\Psi_f|| [c_p^{\dagger}\tilde{c}_n]_K || \Psi_i)$ represent the one-body transition densities (OBTDs), which are model-dependent. In contrast, the functions ${^{V/A}m_{KLS}^{(N)}}(pn)(k_e,m,n,\rho)$ represent the single-particle matrix elements (SPMEs), which are independent of nuclear models for transitions between the initial $(\Psi_i)$ and final $(\Psi_f)$ nuclear states. In this work, the SPMEs are calculated using harmonic oscillator wave functions \cite{Kotila,mst2006}. In Eq. \ref{eq5}, the summation is taken over the single-particle states of protons (p) and neutrons (n). 
In the present calculations, next-to-leading-order (NLO) corrections are also included in the shape factor, leading to an increased number of NMEs (see more about the NLO corrections in Refs. \cite{mika2016,Kotila}).

The shape factor $C(w_e)$ in Eq. (\ref{eq2}) can be broken down into vector, axial-vector, and mixed vector-axial-vector components as a function of electron energy \cite{Kotila,mika2016,Joel_2017,Joel_2017_2,Suhonen1} 
\begin{eqnarray}\label{dcmp}
C(w_e)=g_V^2C_V(w_e)+g_A^2C_A(w_e)+g_Vg_AC_{VA}(w_e).
\end{eqnarray}
By integrating Eq. (\ref{dcmp}) over the electron energy, we get an expression for the integrated shape function $\tilde{C}$, which corresponds  to Eq. (\ref{tc}) 
\begin{eqnarray}\label{intc}
\tilde{C}=g_V^2\tilde{C}_V+g_A^2\tilde{C}_A+g_Vg_A\tilde{C}_{VA}.
\end{eqnarray}

In Eq. (\ref{dcmp}) the shape factors $C_i$ are functions of the electron energy, while the integrated shape factors $\tilde{C_i}$ in Eq. (\ref{intc}) do not depend on the electron energy or simply we can say that the  $\tilde{C_i}$ are just constant numbers.
 
%%%%%%%%%%%%%%%%
\begin{table*}
		\centering
		
		\caption{\label{logft}The calculated  log$ft$ values for the  forbidden non-unique $\beta^{-}$ transitions from the shell model, when s-NMEs are set to zero, are labeled as ``log$ft$(SM)” and after constraining the s-NMEs from CVC theory, they are labeled as ``log$ft$(SM+CVC)”, with coupling constants $g_{A}$=1.00, $g_{A}$=1.27. The ground-state to ground-state Q-values are taken from AME2020 \cite{AME2020}. }

			\begin{tabular}{lccccccccc }
            \hline\hline
				  \multicolumn{1}{c}{~Transition} &K& Q  & BR& Expt.&\multicolumn{2}{c}{log$ft$(SM)}    & \multicolumn{2}{c}{log$ft$ (SM+CVC)} &Interaction\T\B\\%\hline\T\B 

				\cline{1-10}
			 &  &~(keV)& $(\%)$ &  &\multicolumn{2}{c}{$g_{A}$}&\multicolumn{2}{c}{$g_{A}$} & \T\B\\
             \cline{6-7} \cline{8-9}
               &&  &  &&1.00&1.27&1.00& 1.27 & \T\B\\\hline\T\B \\
$^{93}$Zr$(5/2^+) \rightarrow ^{93}$Nb($9/2^+$)  & 2nd   & 90.8(15) & 27 & 12.10(10)&14.765 & 13.497 & 11.847 & 11.752&GWBXG \T\B\\
$^{135}$Cs$(7/2^+) \rightarrow $$^{135}$Ba($3/2^+$) & 2nd  &268.70(29) & 100.0 &13.48(6)&   11.812 &  11.703 &12.935&12.809&SN100PN  \T\B\\
$^{85}$Br$(3/2^-) \rightarrow ^{85}$Kr($9/2^+$)  & 3rd  & 2905(4) & $-$  &$-$ & 13.625  &  13.776  &13.518&13.378&GWBXG\T\B\\
$^{87}$Rb$(3/2^-) \rightarrow $$^{87}$Sr($9/2^+$) & 3rd    & 282.275(6) & 100.0 & 17.514(7)&16.637 & 16.811  &16.430&16.311&GWBXG\T\B\\
$^{97}$Zr$(1/2^+) \rightarrow ^{97}$Nb($9/2^+$) & 4th  & 2666(4) &  $-$  &$-$  & 18.426 & 18.205 &17.345 &17.188&GWBXG \T\B\\
$^{101}$Mo$(1/2^+) \rightarrow ^{101}$Tc($9/2^+$) & 4th   & 2825(24)&  $-$  &$-$  &18.585  & 18.538& 19.164&19.077&GWBXG\T\B\\
$^{115}$Cd$(1/2^+) \rightarrow ^{115}$In($9/2^+$)  & 4th  & 1451.9(7) & $-$  &$-$ & 20.056   &20.265 &19.569 &19.452& G-matrix\T\B\\
$^{117}$Cd$(1/2^+) \rightarrow ^{117}$In($9/2^+$)  & 4th & 2525(5) &$-$  &$-$  & 18.834   & 18.921 &17.863 & 17.713& G-matrix\T\B\\
$^{119}$In$(9/2^+) \rightarrow ^{119}$Sn($1/2^+$) & 4th  & 2366(7) & $-$  &$-$ &  20.004  & 19.944& 20.857&20.770& SNET\T\B\\
$^{123}$Sn$(11/2^-)\rightarrow ^{123}$Sb($1/2^+$) & 5th  & 1408.2(24) & $-$  &$-$ & 26.938  & 27.119 & 26.742 &26.616& SN100PN\T\B\\

\hline
\hline

\end{tabular}
%\end{ruledtabular}
\end{table*}

%%%%%%%%%%%%%%%%%%%%%%
 \subsection{Nuclear Shell Model Calculations} \label{Hamiltonian}
The wave functions of parent and daughter nuclei are calculated using the nuclear shell model. To calculate the OBTDs, we have used the nuclear shell model code KSHELL \cite{KSHELL}. We have used different effective interactions  like GWBXG \cite{gwbxg1,gwbxg2}, G-matrix \cite{G-matrix}, SNET \cite{snet1,snet2}, SN100PN \cite{sn100pn1,sn100pn2} for different model spaces in the SM calculations. The name of the effective interaction is shown in the last column of Table \ref{logft} corresponding to different transitions.

\begin{figure*}
\label{Fig1}

      \includegraphics[scale=0.30]{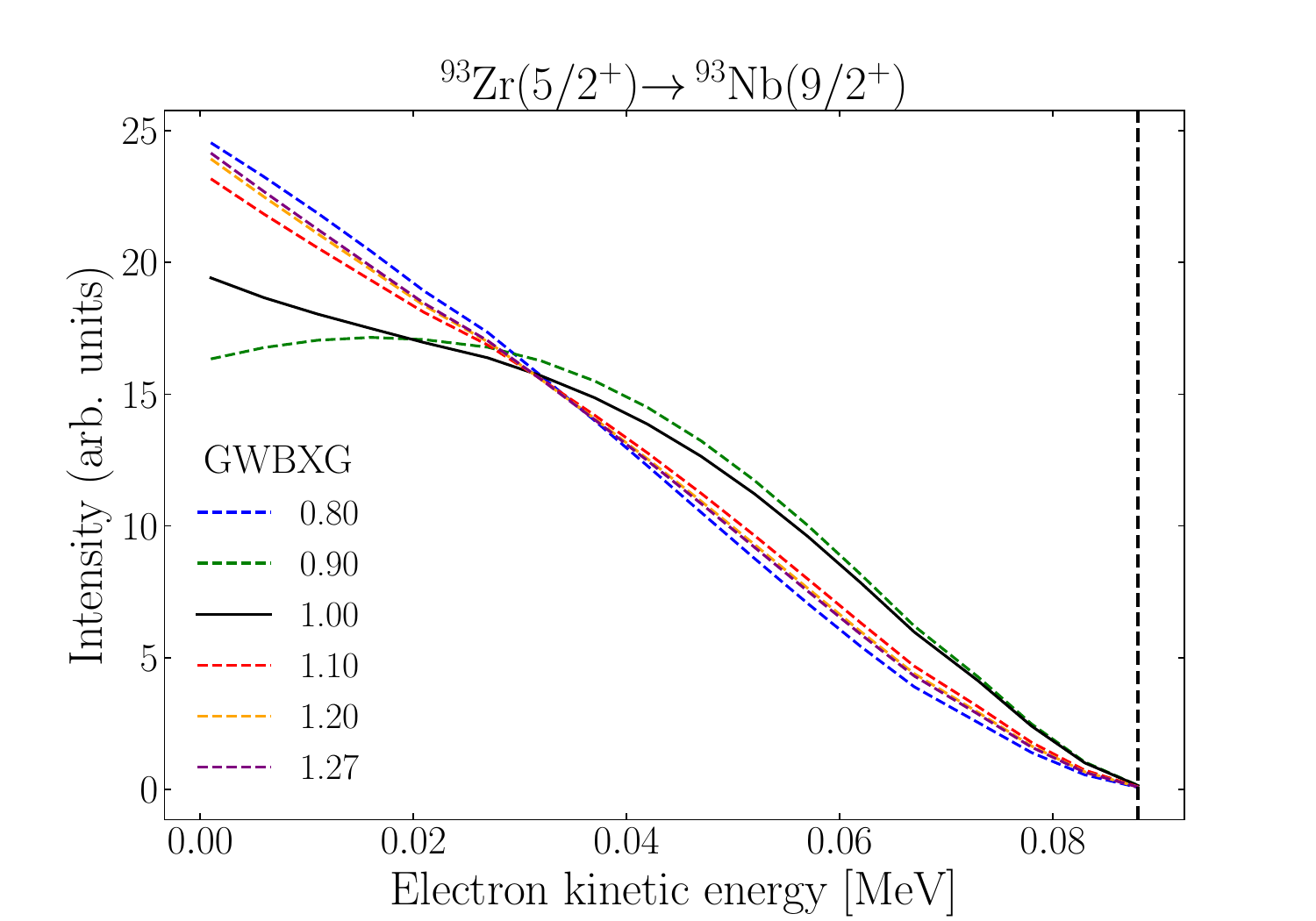}
 \includegraphics[scale=0.30]{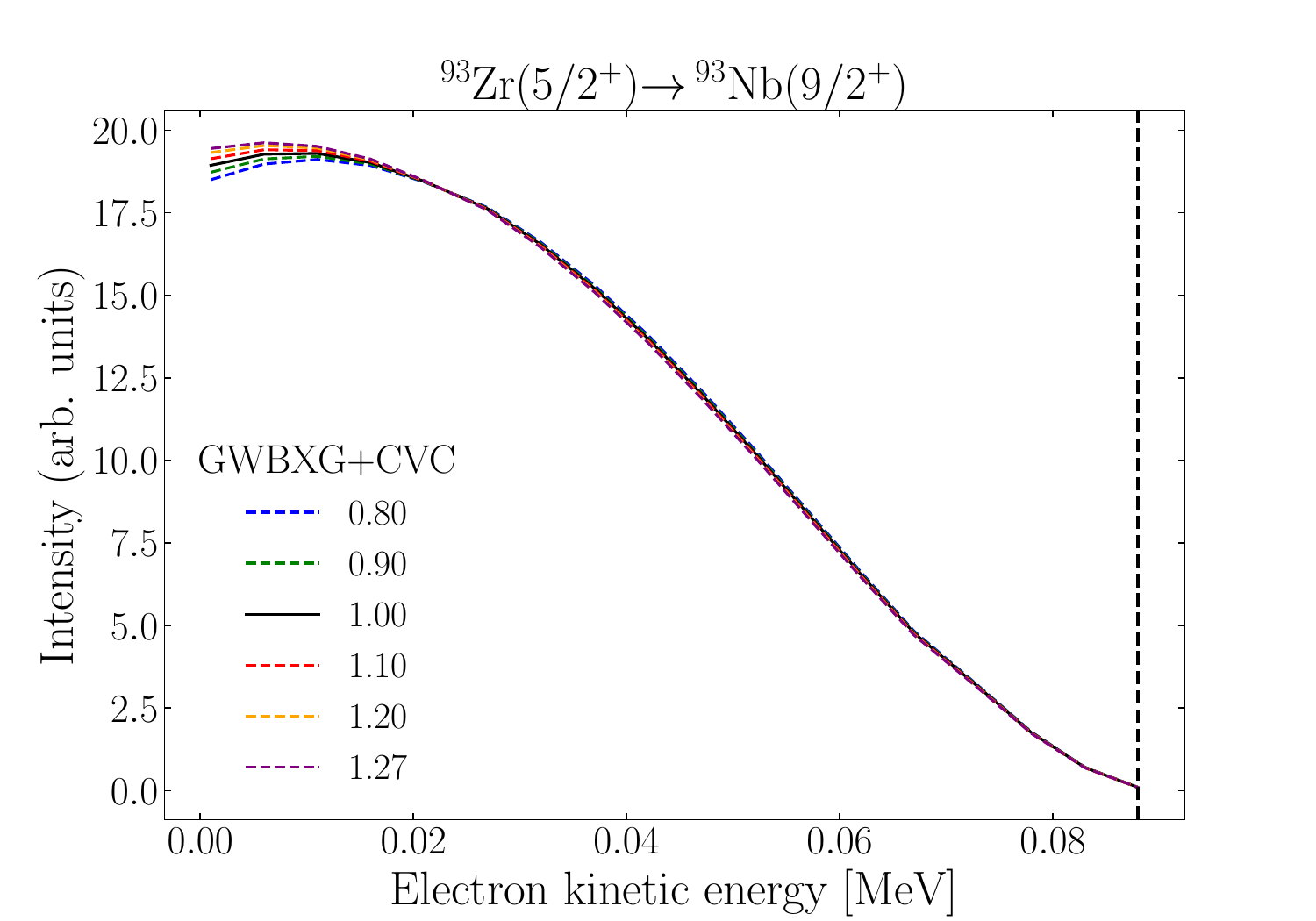}

	\caption{
           The theoretical electron spectra for second forbidden non-unique $\beta^{-}$ transitions of $^{93}$Zr is presented as a function of  electron kinetic energy with $g_V=1.0$ and $g_A=0.80-1.27$. The dashed vertical lines mark the end-point energy for forbidden decay. Each spectrum is normalized such that the area under the curve equals unity.\label{Fig1}}
\end{figure*}

\begin{figure*}
 \includegraphics[scale=0.30]{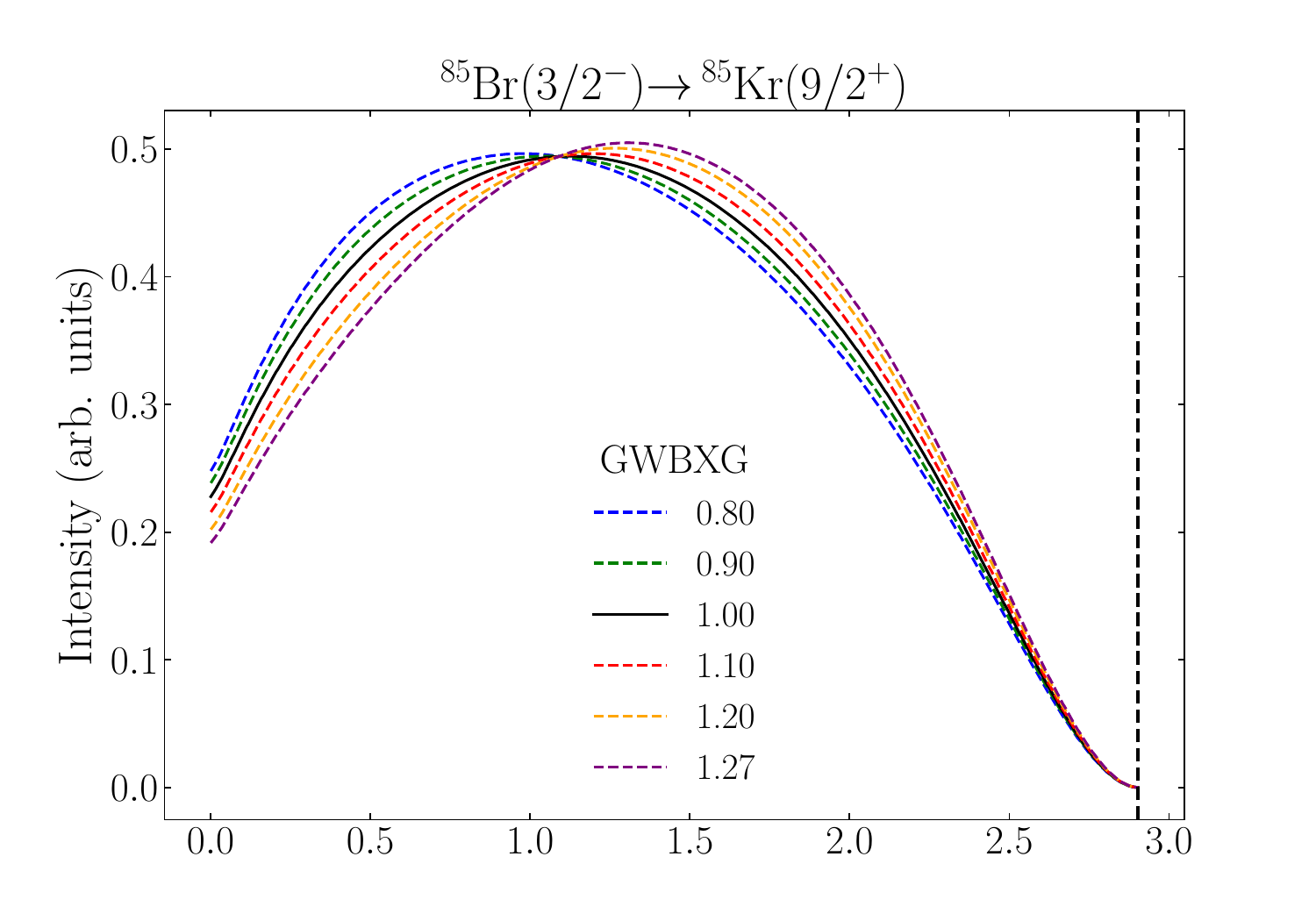}
 \includegraphics[scale=0.30]{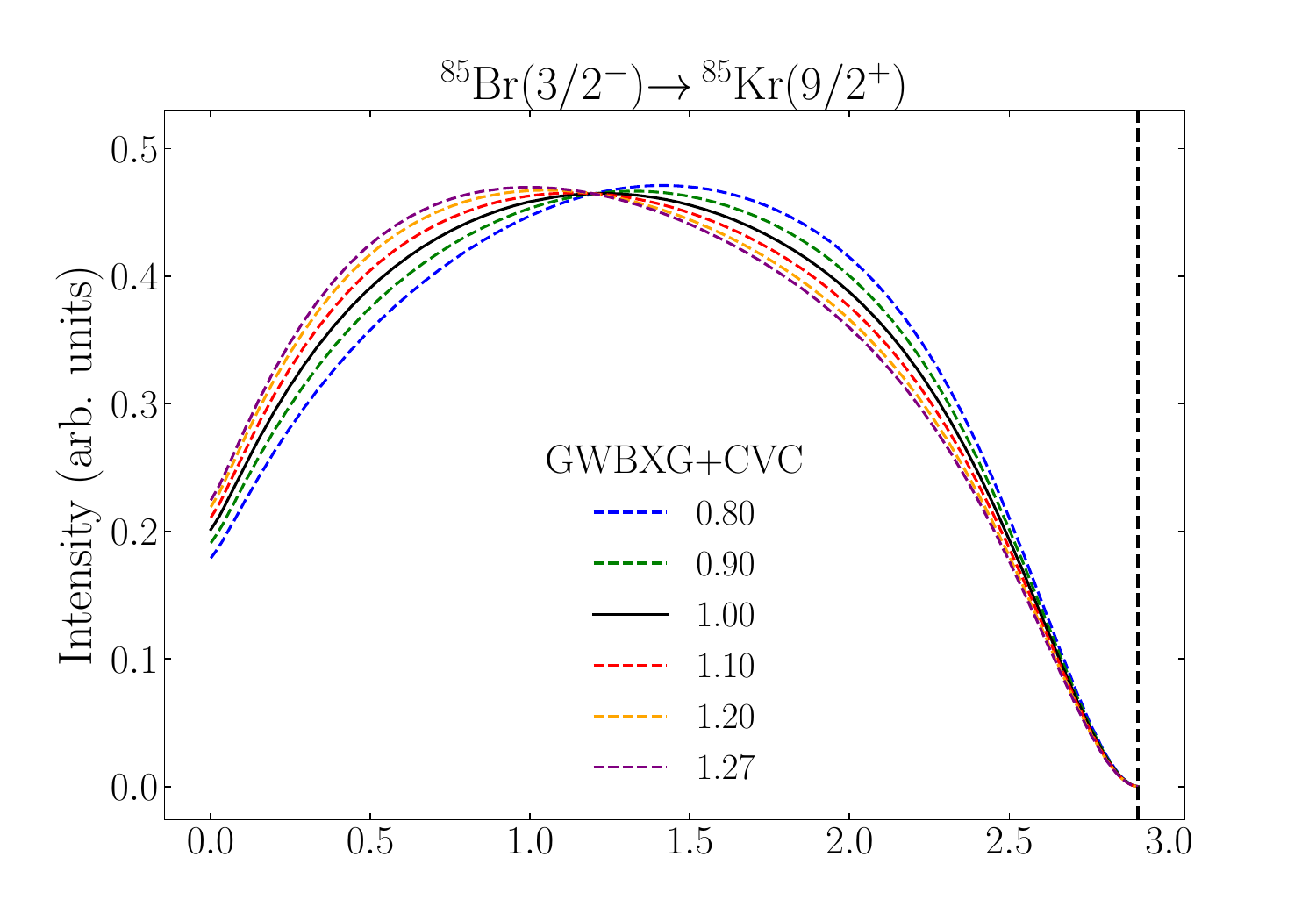}
    \includegraphics[scale=0.30]{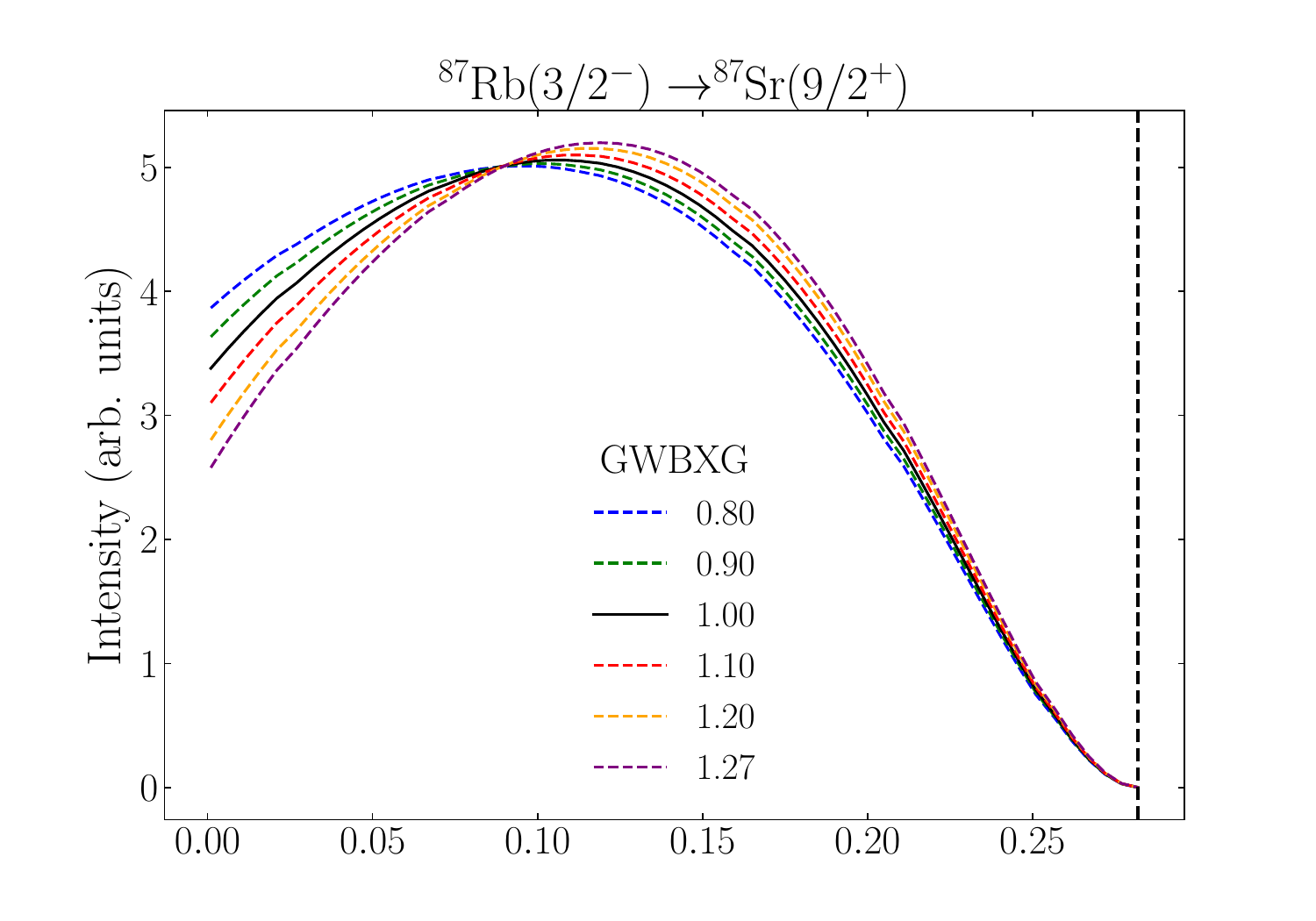} 
\includegraphics[scale=0.30]{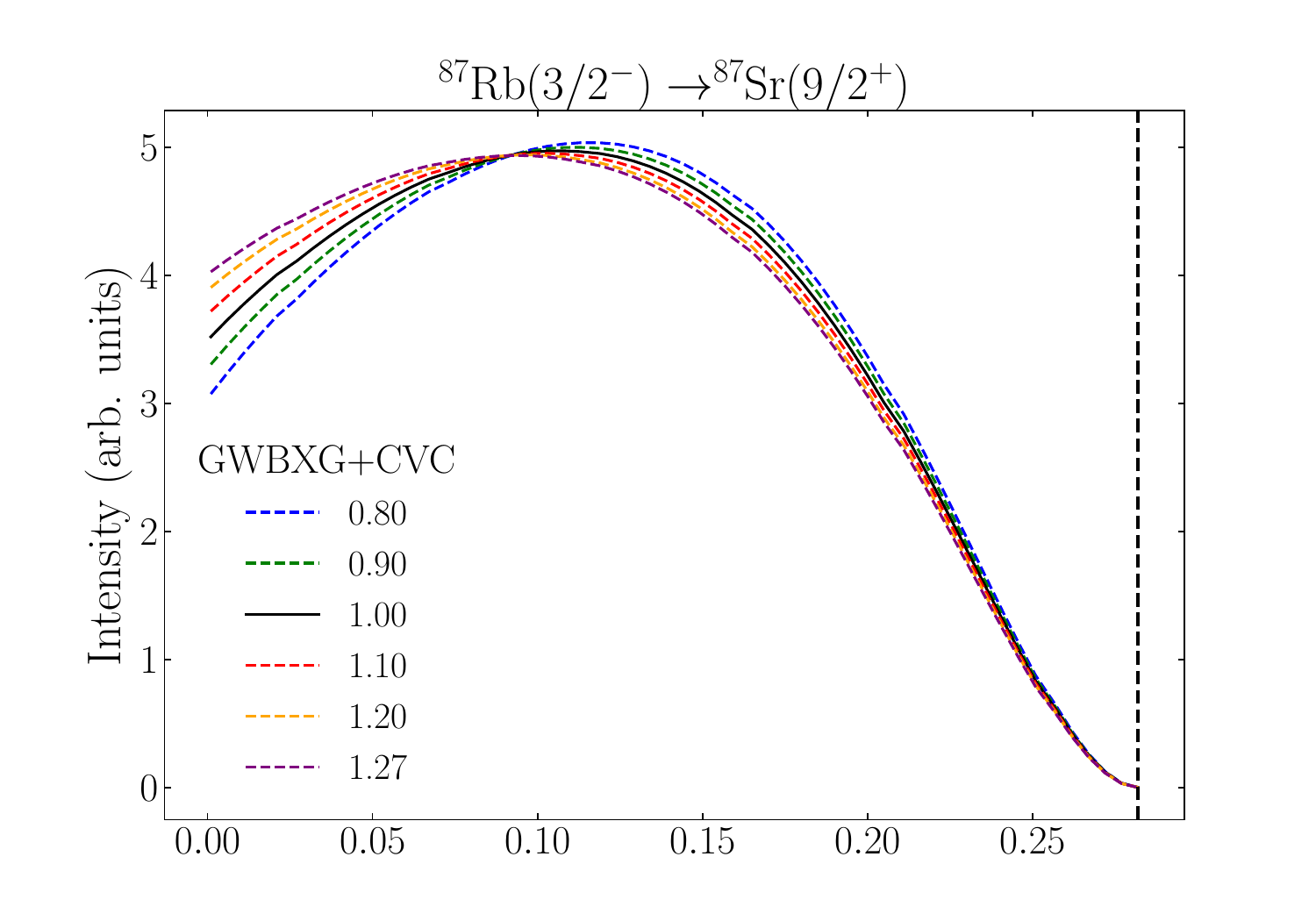} 
\includegraphics[scale=0.30]{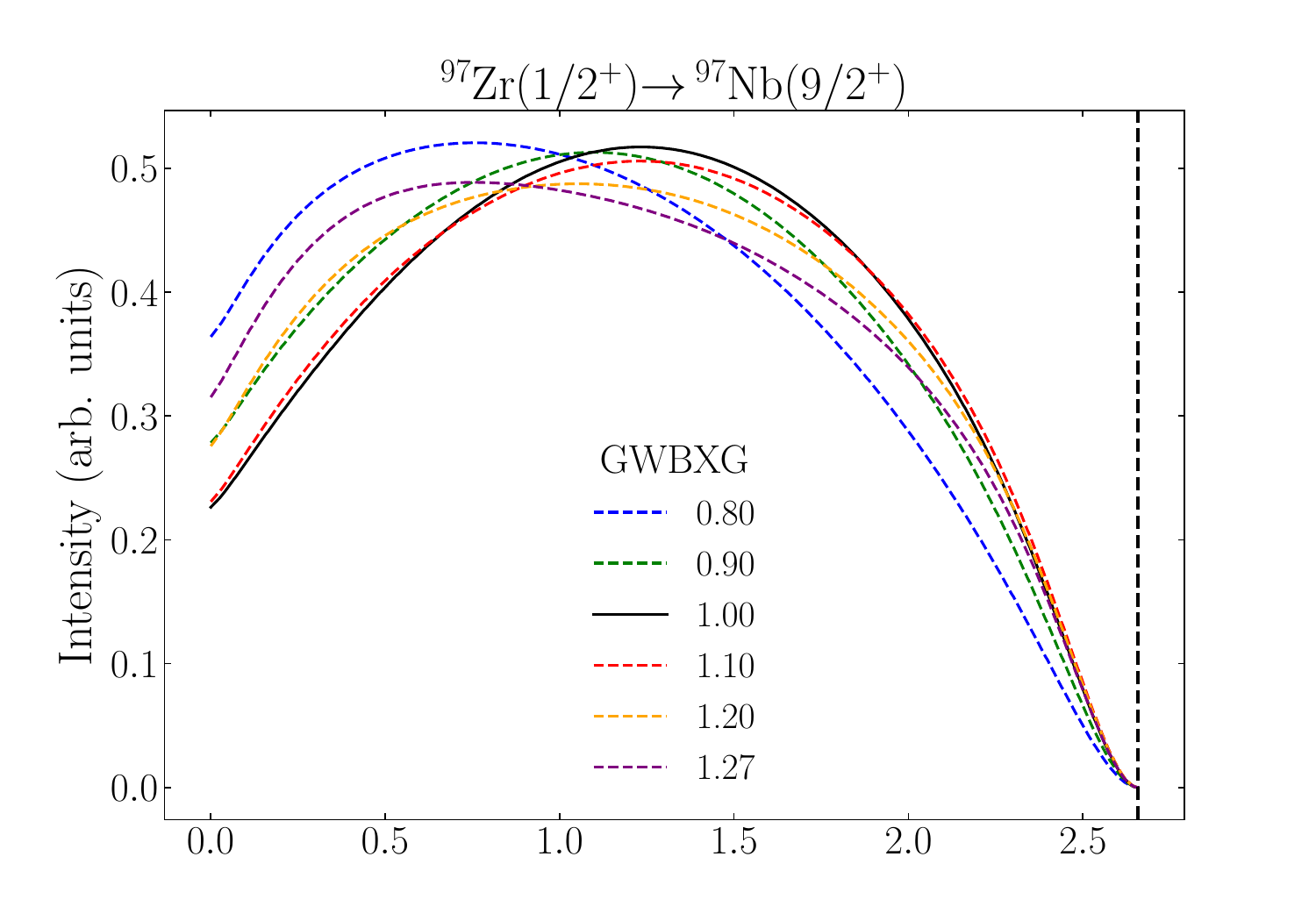}
 \includegraphics[scale=0.30]{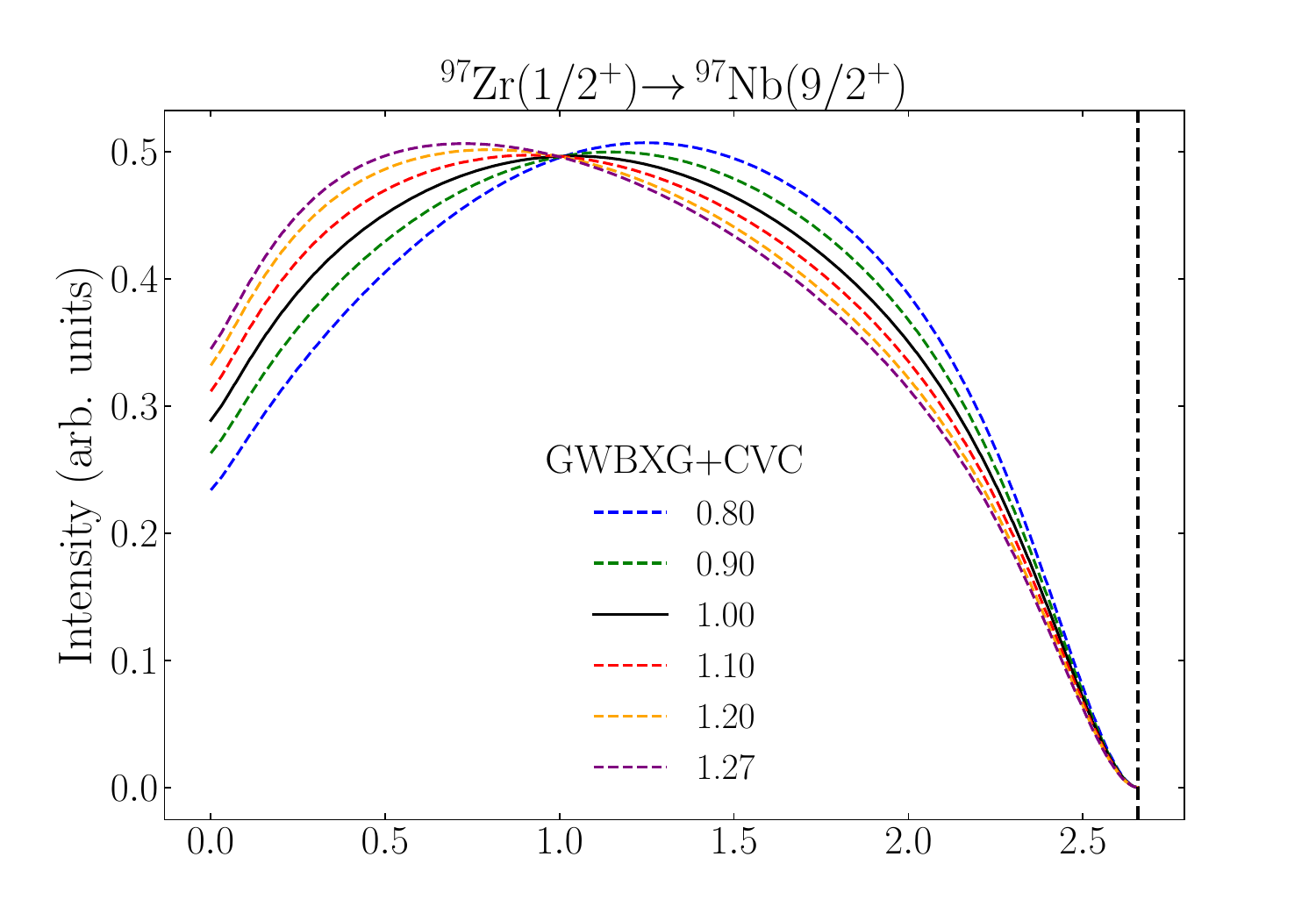}
   \includegraphics[scale=0.30]{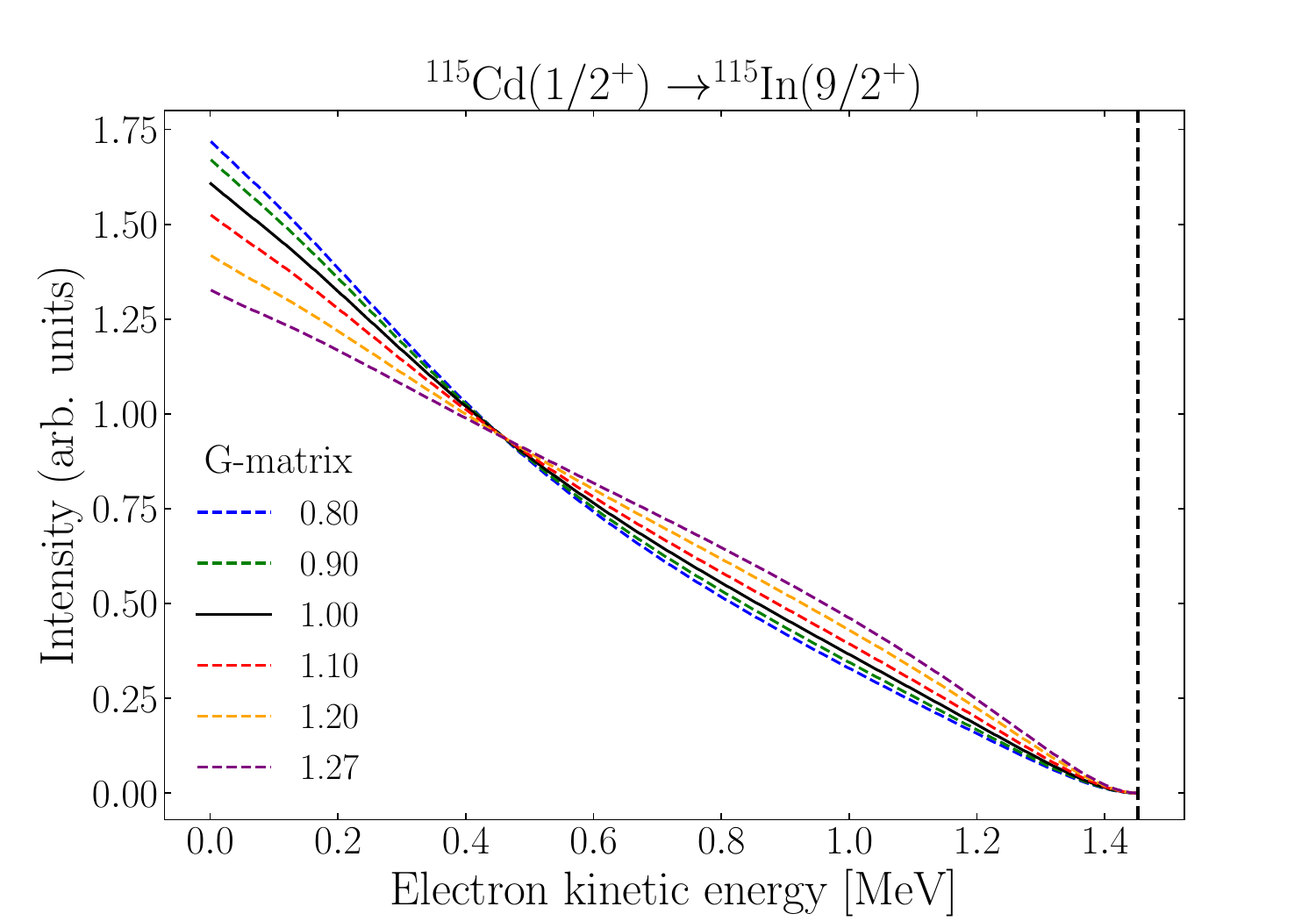}
  \includegraphics[scale=0.30]{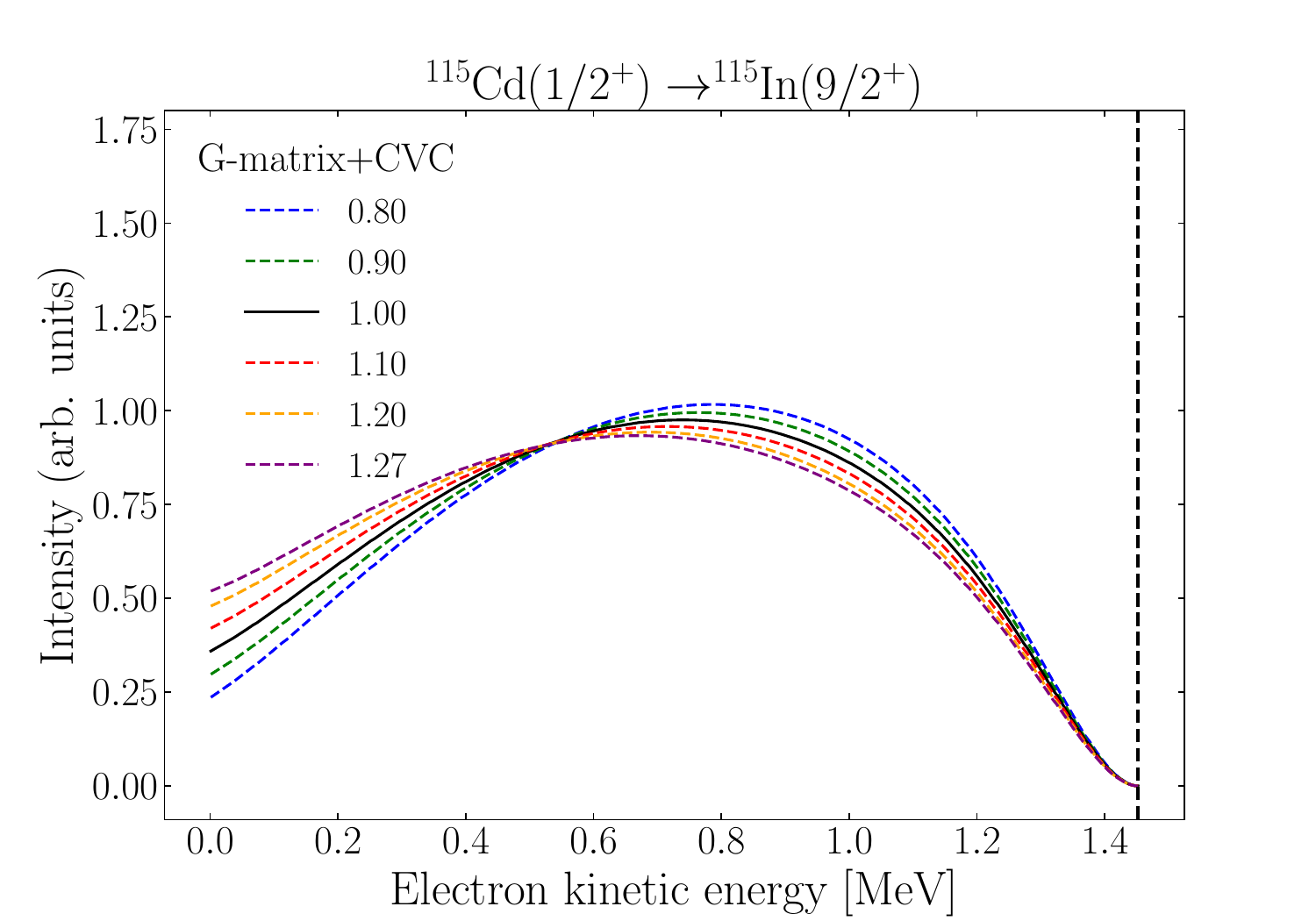}

			\caption{Same as Fig. \ref{Fig1}, but for the third forbidden non-unique $\beta^{-}$ transitions of $^{85}$Br and $^{87}$Rb, and fourth forbidden non-unique transitions of $^{97}$Zr and $^{115}$Cd.\label{Fig2}}
\end{figure*}

\begin{figure*}
  \includegraphics[scale=0.30]{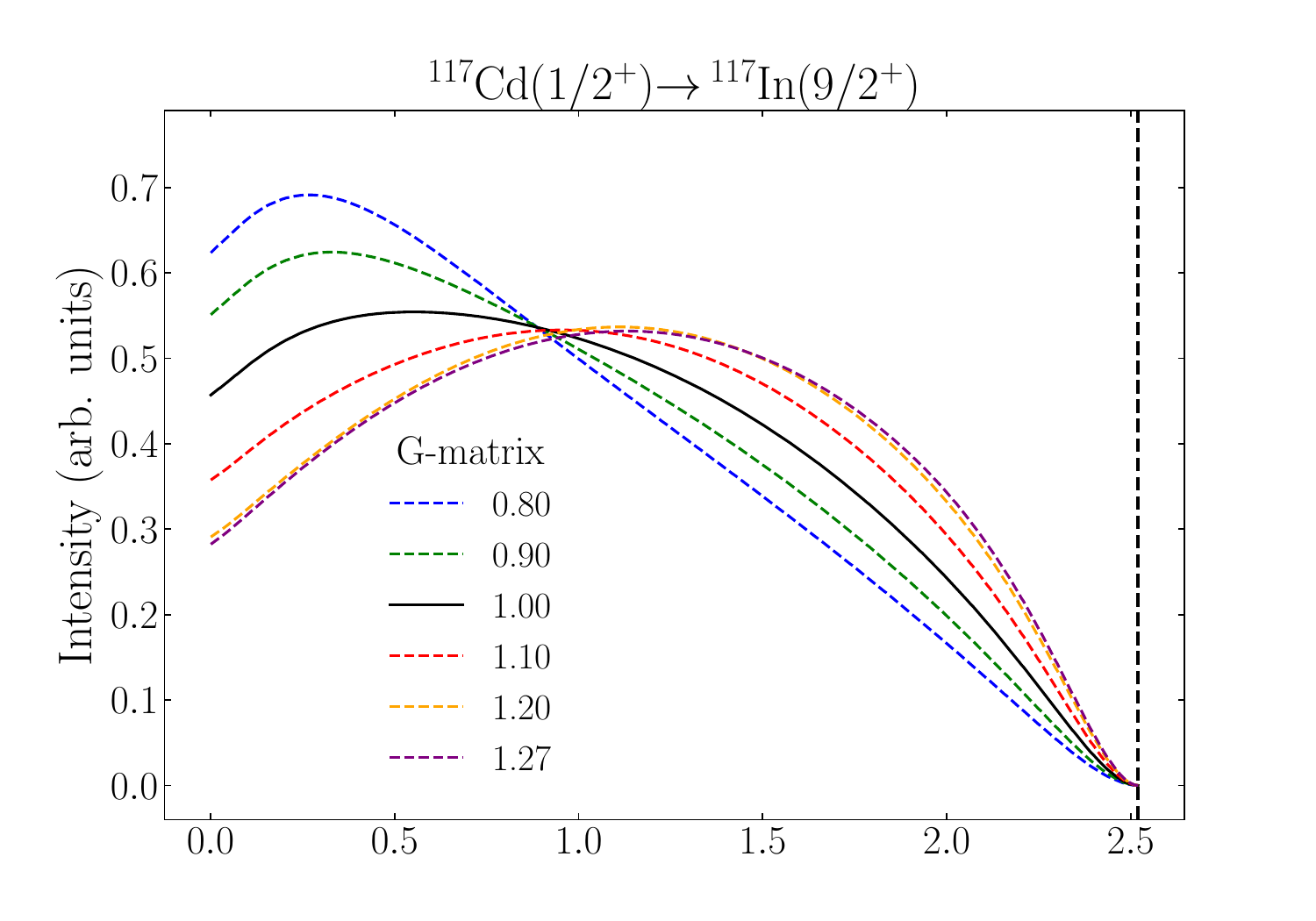}
     \includegraphics[scale=0.30]{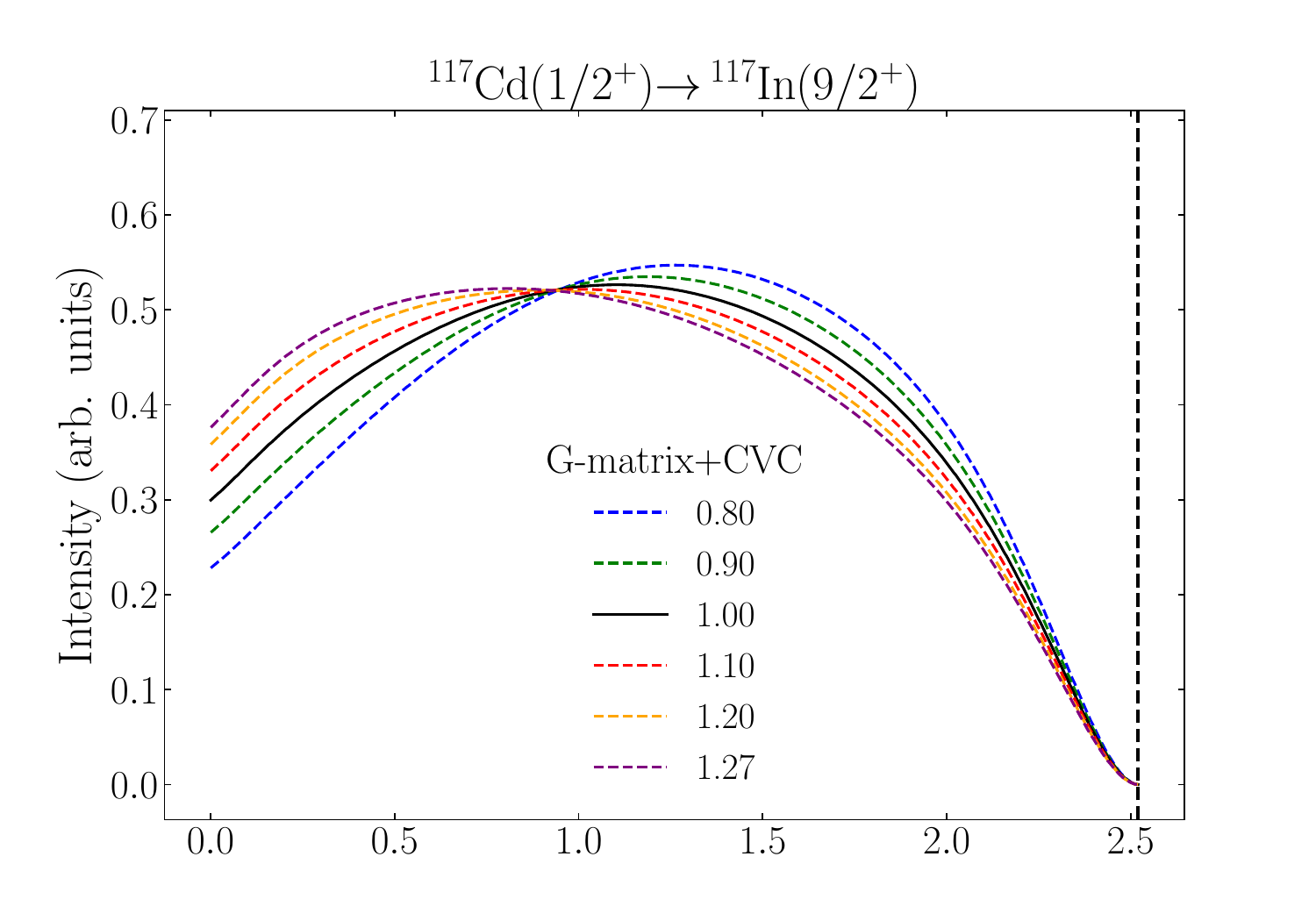}  
  \includegraphics[scale=0.30]{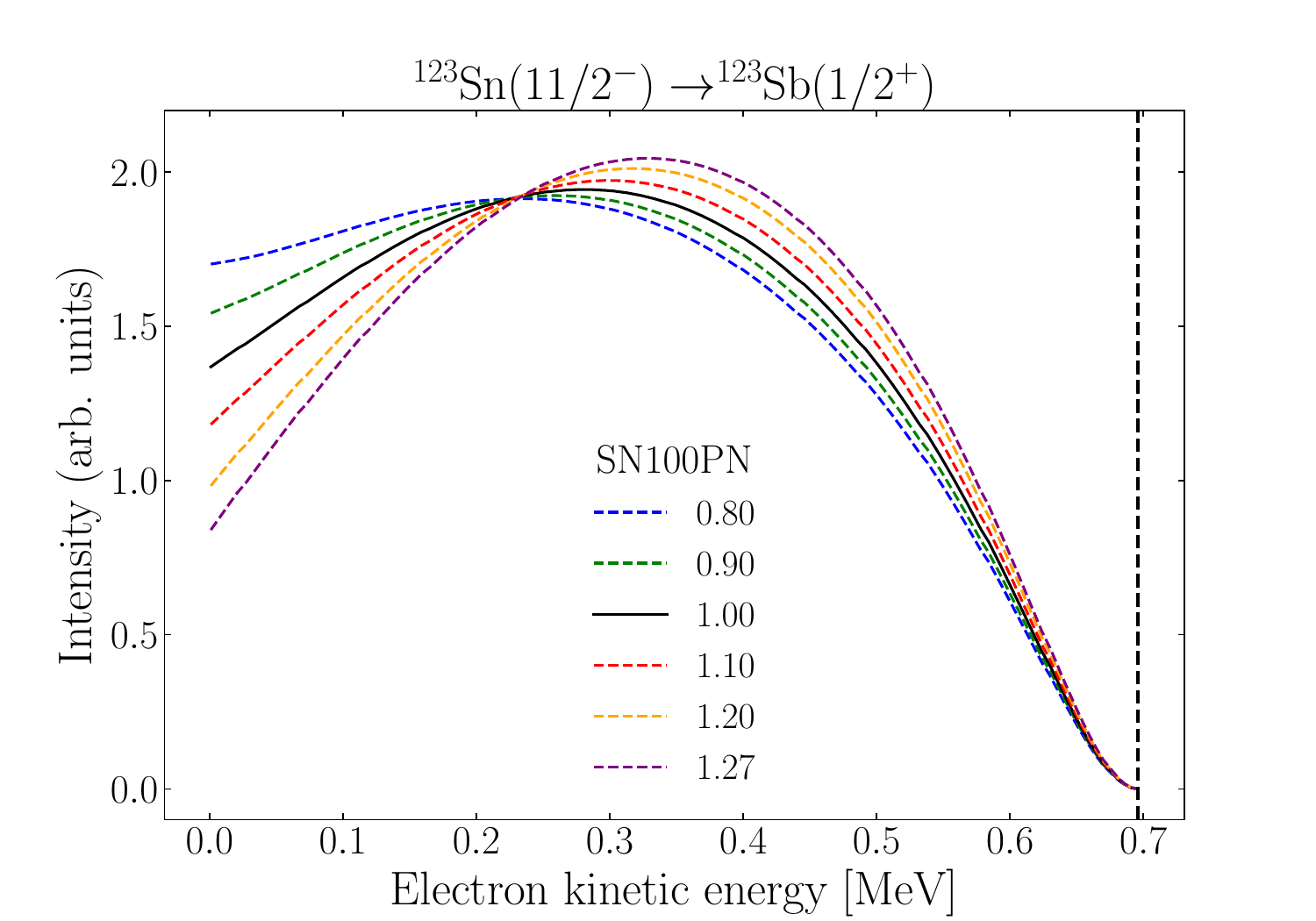}
 \includegraphics[scale=0.30]{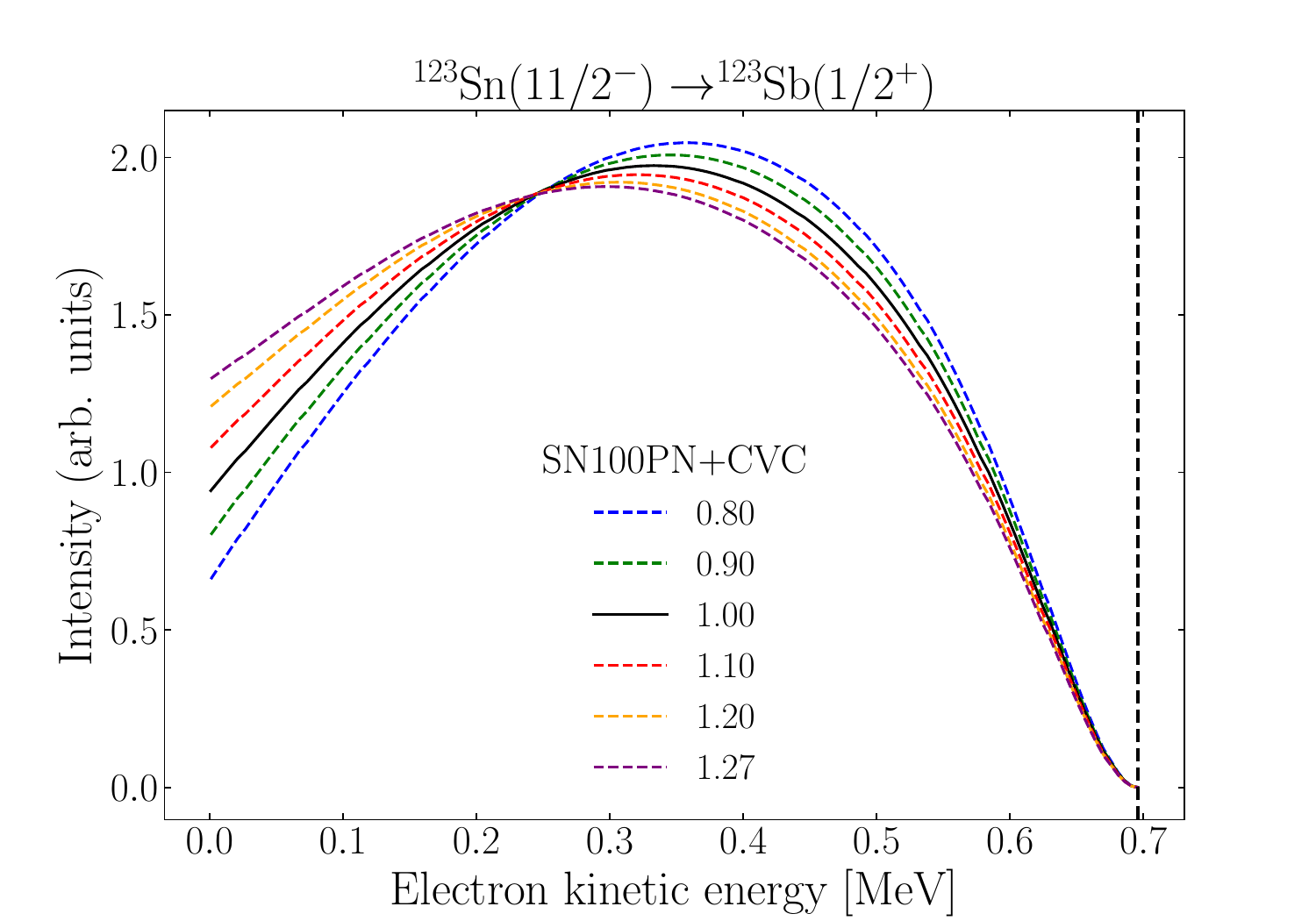}

 \caption{Same as to Fig. \ref{Fig1}, but for the fourth forbidden non-unique $\beta^{-}$ transition of $^{117}$Cd, and fifth forbidden non-unique transition  of $^{123}$Sn.\label{Fig3}}
 \end{figure*}

\begin{figure*}

     \includegraphics[scale=0.30]{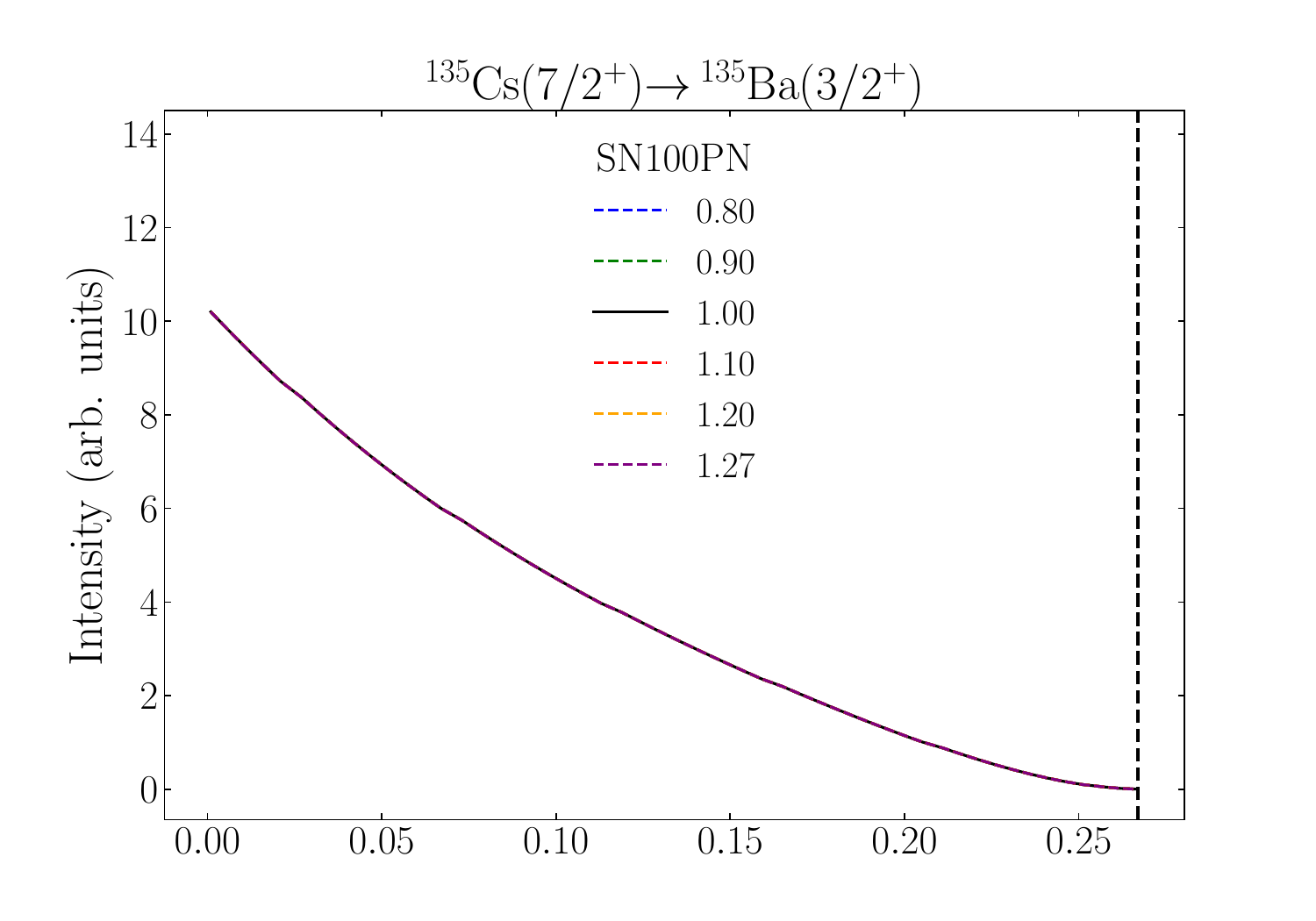}
    \includegraphics[scale=0.30]{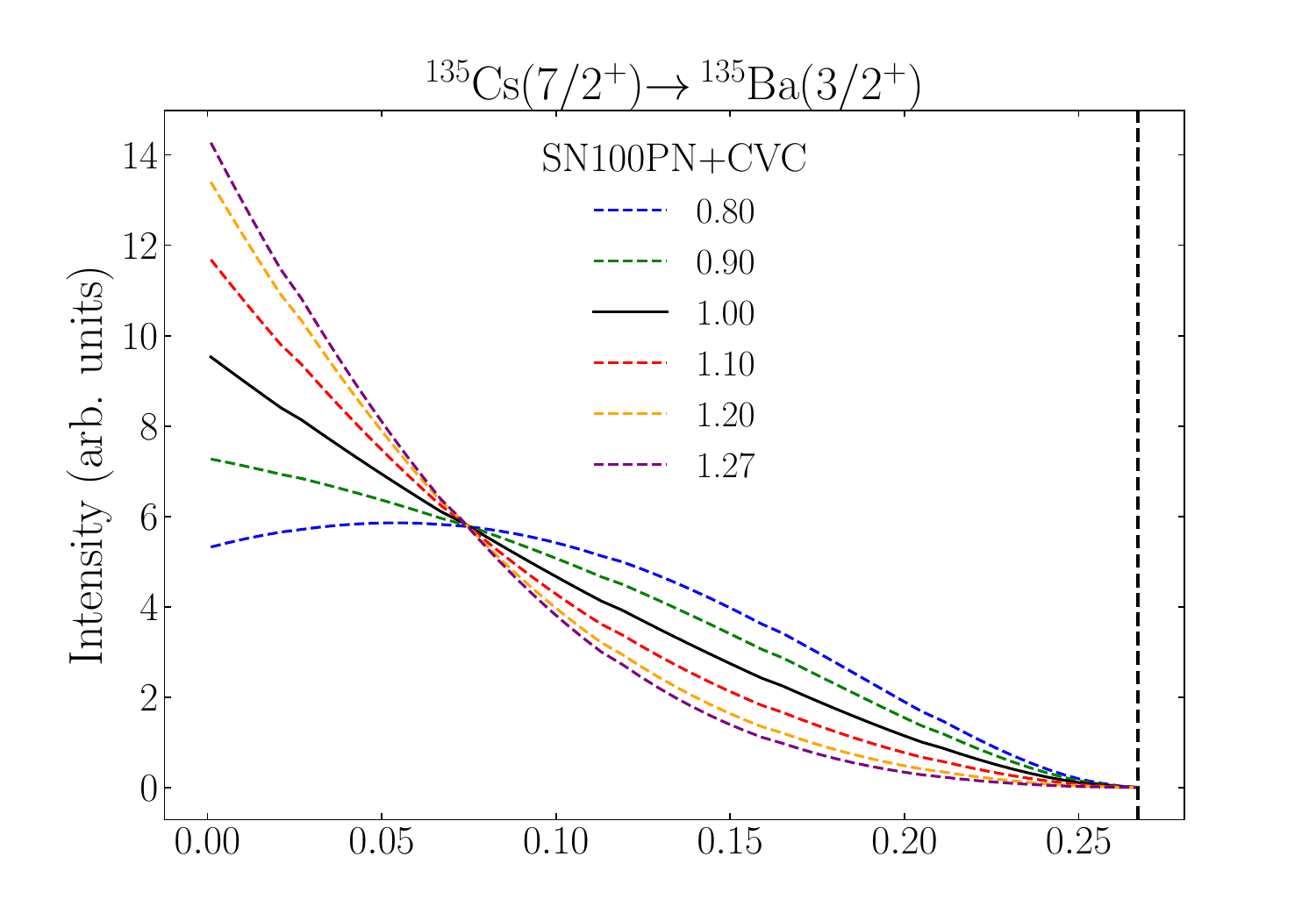} 
 \includegraphics[scale=0.30]{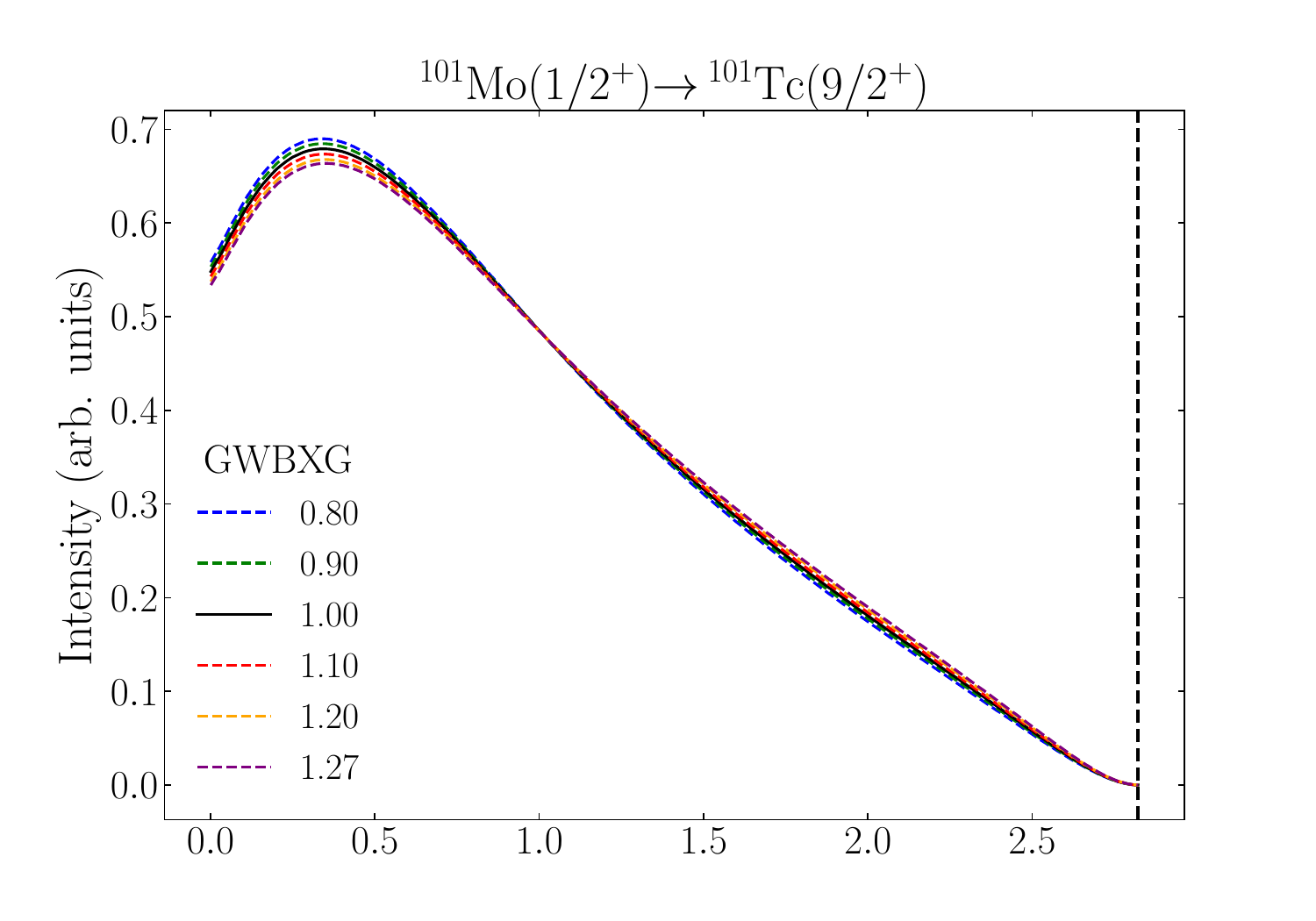}
   \includegraphics[scale=0.30]{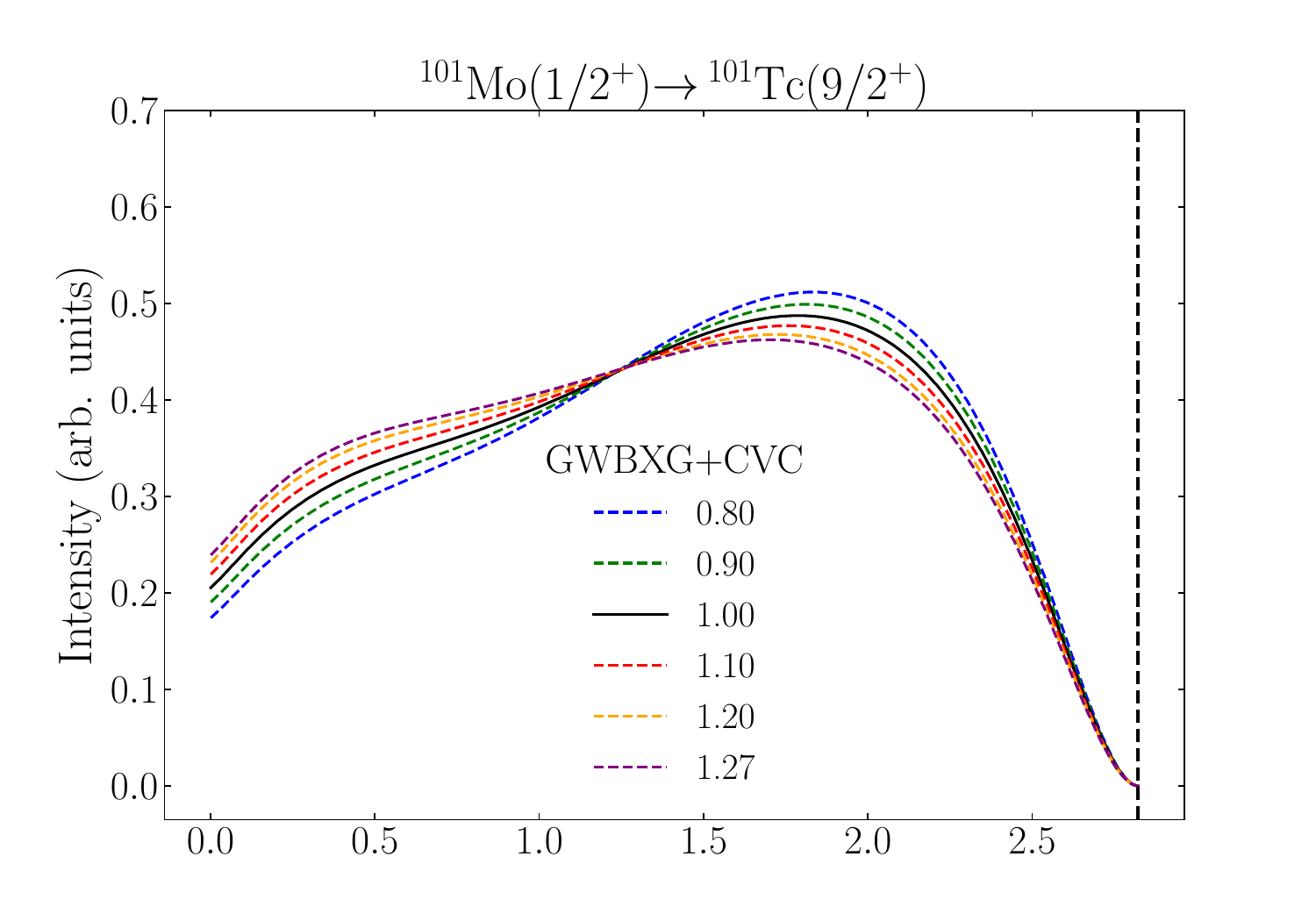}
 \includegraphics[scale=0.30]{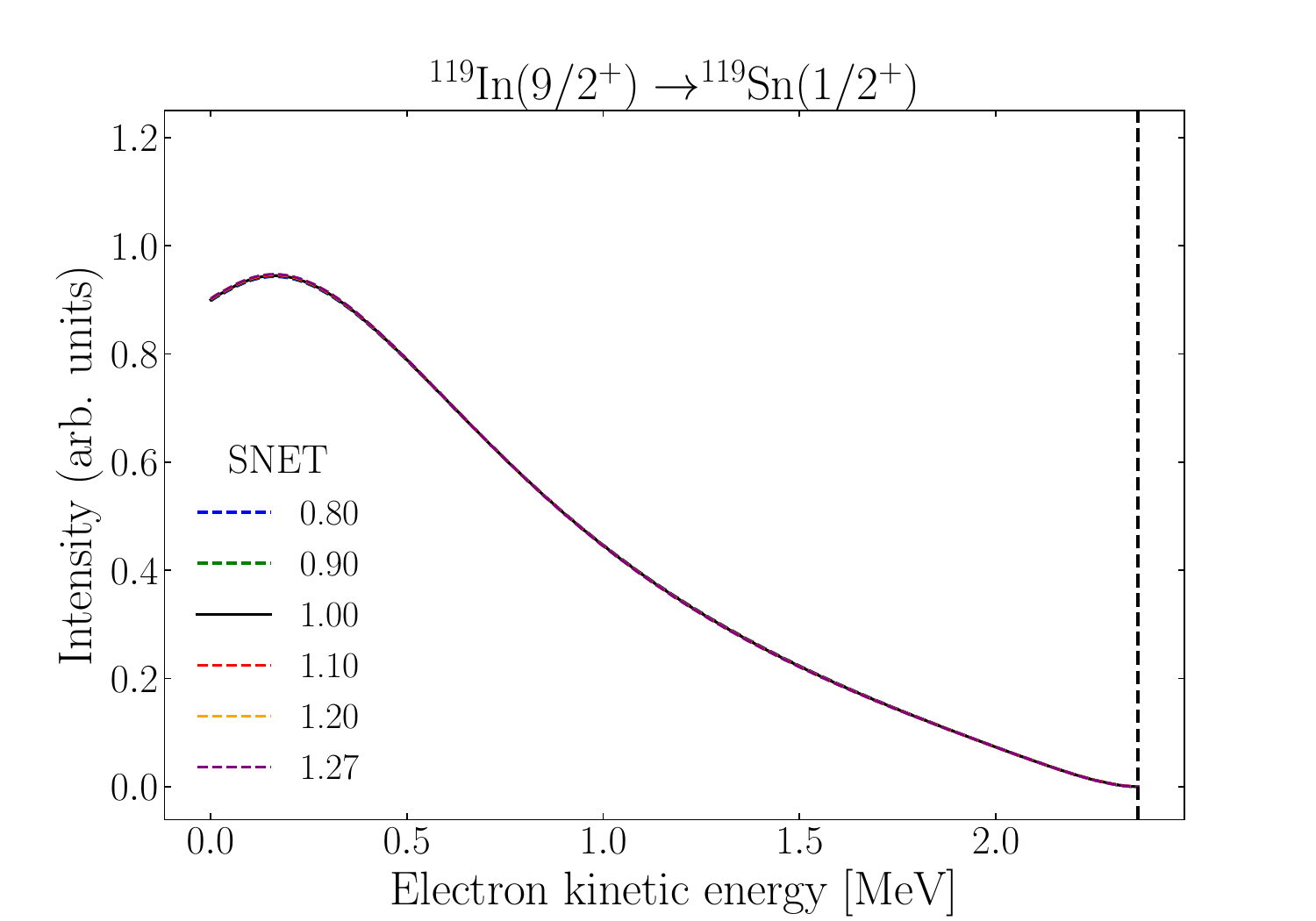}
   \includegraphics[scale=0.30]{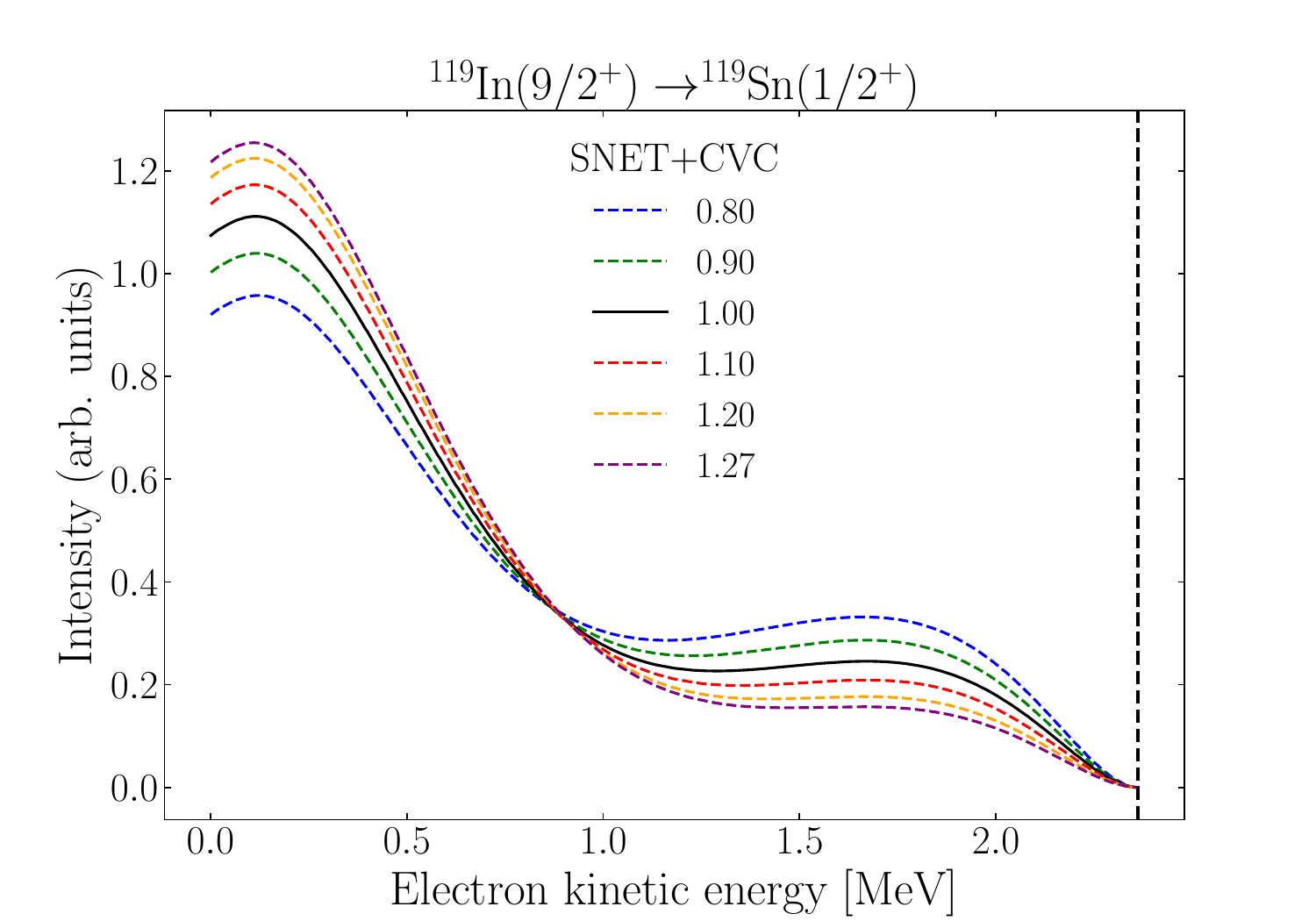}
 
 \caption{Same as Fig. \ref{Fig1}, but for second forbidden non-unique $\beta^{-}$ transitions of $^{135}$Cs, and fourth forbidden non-unique transition of  $^{101}$Mo, and $^{119}$Sn. \label{Fig4}}
 \end{figure*}
 
\begin{table} [!ht]
\leavevmode

\caption{\label{nmes}The calculated l-NME  $^V\mathcal{M}^{(0)}_{KK0}$ and CVC constrained s-NME, $^V\mathcal{M}^{(0)}_{KK-11}$(CVC) for given order of forbiddenness $K$.}
\begin{ruledtabular}
\begin{tabular}{lc}
NMEs/Transition &\multicolumn{1}{c} {$^{93}$Zr$(5/2^+) \rightarrow $$^{93}$Nb($9/2^+$)}
       \T\B\\

\hline 

%$^V\mathcal{M}^{(0)}_{211}$  &        0      \\
$^V\mathcal{M}^{(0)}_{211}$(CVC)  &   -0.2577        \\
%$^V\mathcal{M}^{(0)}_{KK-11}$(Half-life) & 0.1037 & 0.0950   \T\\
$^V\mathcal{M}^{(0)}_{220}$    & -13.8324\\
\hline
NMEs/Transition &\multicolumn{1}{c} {$^{135}$Cs$(7/2^+) \rightarrow $$^{135}$Ba($3/2^+$)} 
       \T\B\\
 
 %\cline{2-3}

\hline 

%$^V\mathcal{M}^{(0)}_{211}$  &        0      \\
$^V\mathcal{M}^{(0)}_{211}$(CVC)  &      0.1612       \\
%$^V\mathcal{M}^{(0)}_{KK-11}$(Half-life) & 0.1037 & 0.0950   \T\\
$^V\mathcal{M}^{(0)}_{220}$    & 7.0161 \\

\hline
NMEs/Transition &\multicolumn{1}{c} {$^{85}$Br$(3/2^-) \rightarrow ^{85}$Kr($9/2^+$)} 
       \T\B\\

\hline 

%$^V\mathcal{M}^{(0)}_{321}$  &        0      \\
$^V\mathcal{M}^{(0)}_{321}$(CVC)  &    -1.3103       \\
%$^V\mathcal{M}^{(0)}_{KK-11}$(Half-life) & 0.1037 & 0.0950   \T\\
$^V\mathcal{M}^{(0)}_{330}$    & -89.3121 \\
\hline
NMEs/Transition &\multicolumn{1}{c} {$^{87}$Rb$(3/2^-) \rightarrow ^{87}$Sr($9/2^+$)} 
       \T\B\\

\hline 

%$^V\mathcal{M}^{(0)}_{321}$  &        0      \\
$^V\mathcal{M}^{(0)}_{321}$(CVC)  &     1.7609         \\
%$^V\mathcal{M}^{(0)}_{KK-11}$(Half-life) & 0.1037 & 0.0950   \T\\
$^V\mathcal{M}^{(0)}_{330}$    & 142.5303 \\

\hline
NMEs/Transition &\multicolumn{1}{c} {$^{97}$Zr$(1/2^+) \rightarrow ^{97}$Nb($9/2^+$)} 
       \T\B\\

\hline 

%$^V\mathcal{M}^{(0)}_{431}$  &        0      \\
$^V\mathcal{M}^{(0)}_{431}$(CVC)  &     9.3769        \\
%$^V\mathcal{M}^{(0)}_{KK-11}$(Half-life) & 0.1037 & 0.0950   \T\\
$^V\mathcal{M}^{(0)}_{440}$    &  791.7846 \\
\hline
NMEs/Transition &\multicolumn{1}{c} {$^{101}$Mo$(1/2^+) \rightarrow ^{101}$Tc($9/2^+$)} 
       \T\B\\

\hline 

%$^V\mathcal{M}^{(0)}_{431}$  &        0      \\
$^V\mathcal{M}^{(0)}_{431}$(CVC)  &     -2.0213      \\
%$^V\mathcal{M}^{(0)}_{KK-11}$(Half-life) & 0.1037 & 0.0950   \T\\
$^V\mathcal{M}^{(0)}_{440}$    & -163.8205 \\
\hline
NMEs/Transition &\multicolumn{1}{c} {$^{115}$Cd$(1/2^+) \rightarrow ^{115}$In($9/2^+$)} 
       \T\B\\

\hline 

$^V\mathcal{M}^{(0)}_{431}$  &        0      \\
$^V\mathcal{M}^{(0)}_{431}$(CVC)  &      -5.0860      \\
%$^V\mathcal{M}^{(0)}_{KK-11}$(Half-life) & 0.1037 & 0.0950   \T\\
$^V\mathcal{M}^{(0)}_{440}$    &  -418.6533\\

\hline 
NMEs/Transition &\multicolumn{1}{c} {$^{117}$Cd$(1/2^+) \rightarrow ^{117}$In($9/2^+$)}   \T\B\\
 \hline     

%$^V\mathcal{M}^{(0)}_{431}$  &        0      \\
$^V\mathcal{M}^{(0)}_{431}$(CVC)  &    -6.9390         \\
%$^V\mathcal{M}^{(0)}_{KK-11}$(Half-life) & 0.1037 & 0.0950   \T\\
$^V\mathcal{M}^{(0)}_{440}$    & -534.3482 \\

\hline 
NMEs/Transition &\multicolumn{1}{c} {$^{119}$In$(9/2^+) \rightarrow ^{119}$Sn($1/2^+$)}   \T\B\\
     
\hline 
%$^V\mathcal{M}^{(0)}_{431}$  &        0      \\
$^V\mathcal{M}^{(0)}_{431}$(CVC)  &   0.5265         \\
%$^V\mathcal{M}^{(0)}_{KK-11}$(Half-life) & 0.1037 & 0.0950   \T\\
$^V\mathcal{M}^{(0)}_{440}$    & 40.4356 \\

\hline 
NMEs/Transition &\multicolumn{1}{c} {$^{123}$Sn$(11/2^-) \rightarrow ^{123}$Sb($1/2^+$)}   \T\B\\
     
\hline 
%$^V\mathcal{M}^{(0)}_{541}$  &        0      \\
$^V\mathcal{M}^{(0)}_{541}$(CVC)  &   -12.227         \\
%$^V\mathcal{M}^{(0)}_{KK-11}$(Half-life) & 0.1037 & 0.0950   \T\\
$^V\mathcal{M}^{(0)}_{550}$    & -1290.0080 \\

\end{tabular}
\end{ruledtabular}
%\end{small}
\end{table}

\begin{table*}

\caption{\label{decomposition} The dimensionless integrated shape factors $\tilde{C}$ for the studied transitions,  and their decompositions to vector $\tilde{C}_V$, axial-vector $\tilde{C}_A$, and vector-axial-vector $\tilde{C}_{VA}$ parts with  $g_V=g_A=1.0$. The results after constraining the s-NMEs are labeled as ``SM+CVC".}
\begin{ruledtabular}
\begin{tabular}{lcccc} \\

Transition &$\tilde{C}_V$ &$\tilde{C}_A$ & $\tilde{C}_{VA}$ &  $\tilde{C}$ \\
\hline

$^{93}$Zr$(5/2^+) \rightarrow $$^{93}$Nb($9/2^+$)\\

 SM& $1.00103\times 10^{-11}$ &$ 1.15633\times 10^{-11}$  & $-2.15004\times 10^{-11}$ & $ 7.31903\times 10^{-14}$  \\

SM+CVC &  $2.03842\times 10^{-11}$ & $1.15633\times 10^{-11}$  & $2.86341\times 10^{-11}$ & $6.05817\times 10^{-11}$\\
\hline

$^{135}$Cs$(7/2^+) \rightarrow $$^{135}$Ba($3/2^+$)\\
SM & $1.05786\times 10^{-9}$ &$1.01622\times 10^{-9}$  & $2.07190\times 10^{-9}$ &  $4.14598\times 10^{-9}$ \\

SM+CVC &  $1.21435 \times 10^{-9}$& $1.01622  \times 10^{-9}$& $-1.91878\times 10^{-9}$ &  $3.11787\times 10^{-10}$ \\
\hline
$^{85}$Br$(3/2^-) \rightarrow ^{85}$Kr($9/2^+$) \\

SM& $5.00605\times 10^{-7}$ & $1.04551\times 10^{-7}$ &$-4.28820\times 10^{-7}$ & $1.76336\times 10^{-7}$ \\
SM+CVC & $4.26342\times 10^{-8}$& $ 1.04551\times 10^{-7}$&  $7.86317\times 10^{-8} $ &  $2.25817\times 10^{-7} $\\
\hline

$^{87}$Rb$(3/2^-) \rightarrow ^{87}$Sr($9/2^+$) \\

SM & $8.80885\times 10^{-14}$ &$1.59705\times 10^{-14}$  & $-7.34945\times 10^{-14}$ & $3.05644\times 10^{-14}$  \\
SM+CVC & $1.19390\times 10^{-14}$ & $1.59705\times 10^{-14}$& $2.13381\times 10^{-14}$ & $4.92477\times 10^{-14}$\\
\hline
$^{97}$Zr$(1/2^+) \rightarrow ^{97}$Nb($9/2^+$) \\

SM  & $1.80481\times 10^{-11}$ & $1.68290\times 10^{-11}$  & $ -3.25912\times 10^{-11}$ & $ 2.28597\times 10^{-12}$  \\

SM+CVC  & $ 4.62979\times 10^{-12}$ & $1.68290\times 10^{-11}$  & $6.11628\times 10^{-12}$ & $2.75751\times 10^{-11}$ \\
\hline
$^{101}$Mo$(1/2^+) \rightarrow ^{101}$Tc($9/2^+$) \\
SM & $1.58876\times 10^{-12}$ & $ 2.45600\times 10^{-13}$ & $ 3.69890\times 10^{-13} $& $2.20425\times 10^{-12}$  \\

SM+CVC &  $4.12088\times 10^{-13}$ & $2.45600\times 10^{-13}$ & $-7.71379\times 10^{-14} $& $5.80550\times 10^{-13}$ \\

\hline
$^{115}$Cd$(1/2^+) \rightarrow ^{115}$In($9/2^+$) \\
SM & $2.49942\times 10^{-14}$ & $7.69019\times 10^{-15}$ & $-2.63806\times 10^{-14}$ & $6.30379\times 10^{-15}$  \\
SM+CVC &  $6.90690\times 10^{-15}$ & $7.69019\times 10^{-15}$ & $4.74511\times 10^{-15}$ & $1.93422\times 10^{-14}$\\
\hline
$^{117}$Cd$(1/2^+) \rightarrow ^{117}$In($9/2^+$) \\
SM&  $8.37700 \times 10^{-12}$ & $5.32112\times 10^{-12}$ & $-1.27295\times 10^{-11}$& $9.68629\times 10^{-13}$\\
SM+CVC &  $1.91622\times 10^{-12}$ & $5.32112\times 10^{-12}$ & $1.81894\times 10^{-12}$& $9.05628\times 10^{-12}$  \\

\hline

$^{119}$In$(9/2^+) \rightarrow ^{119}$Sn($1/2^+$) \\
SM&$2.82952\times 10^{-14}$  & $3.64590\times 10^{-15}$ & $2.01459\times 10^{-14}$ & $5.20870\times 10^{-14}$  \\
SM+CVC&  $5.96564\times 10^{-15}$  & $3.64590\times 10^{-15}$ & $-2.30828\times 10^{-15}$ & $7.30327\times 10^{-15}$ \\
\hline
$^{123}$Sn$(11/2^-)\rightarrow ^{123}$Sb($1/2^+$) \\
SM & $1.81951\times 10^{-22}$ & $3.78309\times 10^{-23}$  & $-1.60598\times 10^{-22}$ & $5.91838\times 10^{-23}$  \\
SM+CVC & $2.53090\times 10^{-23}$ & $3.78309\times 10^{-23}$  & $2.99880\times 10^{-23}$ & $9.31278\times 10^{-23}$ \\

\end{tabular}
\end{ruledtabular}
%\end{small}
\end{table*}

%\textcolor{blue}{About gwbxg effective interaction}
For the decay of $^{93}$Zr, $^{85}$Br, $^{87}$Rb, $^{97}$Zr and  $^{101}$Mo, the GWBXG effective interaction is used. The model space of GWBXG effective interaction consists of
1$f_{5/2}$, 2$p_{3/2}$, 2$p_{1/2}$, 1$g_{9/2}$  proton orbitals and 
2$p_{1/2}$, 1$g_{9/2}$, 1$g_{7/2}$, 2$d_{5/2}$, 2$d_{3/2}$, and 3$s_{1/2}$  neutron orbitals using  $^{68}$Ni as a core, for more details, see Refs. \cite{H7B, Ji1988, Gloeckner1975, Serduke1976}. For $^{85}$Br, $^{87}$Rb and $^{93}$Zr, we  do one neutron excitation from 1$g_{9/2}$ to 1$g_{7/2}$, 2$d_{5/2}$, 2$d_{3/2}$, and 3$s_{1/2}$ orbitals. 
 %\textcolor{blue}{About gmatrix effective interaction}
For the decay of $^{115}$Cd and $^{117}$Cd, we have used the G-matrix effective interaction. The model space has two proton orbitals 1$p_{1/2}$ and 0$g_{9/2}$ and five neutron orbitals 2$d_{5/2}$, 3$s_{1/2}$, 2$d_{3/2}$, 1$g_{7/2}$ with $^{88}$Sr as an inert core \cite{Machleidt2001, G-matrix}. Using G-matrix effective interaction we have not employed any truncation.
%\textcolor{blue}{About snet effective interaction} 
For the decay of $^{119}$In, we have used the SNET effective interaction. The model space of SNET effective interaction has eight  proton orbitals and nine neutron orbitals. In the calculations using SNET effective interaction, we have put a truncation for proton orbitals as the $1f_{5/2}$ $2p_{3/2}$, $2p_{1/2}$ are completely filled and no protons are occupied in the $2d_{5/2}$, $2d_{3/2}$ and $3s_{1/2}$ orbitals while 8-10 protons in $1g_{9/2}$, and 0-2 protons in $1g_{7/2}$ orbital. While the neutron orbitals $1f_{5/2}$ $2p_{3/2}$, $2p_{1/2}$ $1g_{9/2}$ are completely filled and $1g_{7/2}$ $2d_{5/2}$, $2d_{3/2}$ $3s_{1/2}$ are completely open and restricted 0-4 neutrons in $1h_{11/2}$ orbital.
For the decay of $^{135}$Cs and $^{123}$Sn, we have used SN100PN effective interaction. No truncation has been employed using SN100PN effective interaction which has model space $50-82$ having 1$g_{7/2}$, 2$d_{5/2}$, 2$d_{3/2}$,  3$s_{1/2}$, 1$h_{11/2}$, orbitals \cite{sn100pn2, sn100pn1}.

\section{Results and Discussion} \label{result}
We present the electron spectral-shapes of higher forbidden non-unique $\beta^{-}$ transitions in the mass range $A=85-123$ for  $g_{V}= 1.0$ and  $g_{A}= 0.8-1.27$  in Figs. \ref{Fig1}-\ref{Fig4}. The computed small NME (s-NME) from CVC relation is shown in Table \ref{nmes}. Table \ref{logft} presents the log$ft$ values for $g_{V}= 1.0$ and $g_{A}= 1.0$ and $1.27$, and Table \ref{decomposition} shows the decomposition of shape factor  for $g_{V}= g_{A}$=1.0 in the calculations.

\subsection{Nuclear Matrix Elements} \label{NME}
In the SM calculations, the value of s-NME becomes zero because of 0$\hbar \omega$ calculation in the adopted  model space. The s-NME plays a crucial role in the understanding of electron spectral-shapes which are discussed in recent works \cite{joel2021, Anil2020PRC, anil2021}. 
The value of  s-NME can be determined in two ways: (i) using the CVC relation, and (ii) by adjusting s-NME to reproduce the experimental partial half-life.
In the present manuscript, we have used the CVC relation to get the value of s-NME. The CVC theory establishes a framework for connecting  s-NME  with the large vector NME (l-NME) \cite{behrens1982, Behrens1971} and the relation between  s-NME ($^V\mathcal{M}^{(0)}_{KK-11}$ ) and the l-NME ($^V\mathcal{M}^{(0)}_{KK0}$) is given as:

%\begin{figure}%[htbp]
\begin{equation}
\label{eq:CVC-sNME}
   ^V\mathcal{M}_{KK-11}^{(0)} = \left(\frac{\frac{{(-M_n c^2 + M_p c^2 + W_0) \times R}}{{\hbar c}} + \frac{6}{5} \alpha Z}{\sqrt{K (2K + 1)} \times R}\right) \times {}^V\mathcal{M}_{KK0}^{(0)}. \
\end{equation}
% \end{figure}
 
In Eq. \ref{eq:CVC-sNME}, $M_{n}$ and $M_{p}$ are the masses of the neutron and proton, respectively. The nuclear radius is given by $R=1.27A^{1/3}$ in fm. The s-NME which are called relativistic matrix elements, are the product of small and large components of initial and final wave functions of radial Dirac equations and l-NME which are called non-relativistic matrix elements are the product of both small components of the initial and final wave functions, as well as the product of both large components of the same wave functions, of radial Dirac equations, for more details, see Refs. \cite{Gregorio2024,Behrens1971,Sadler, Zhi2013,Dag_thesis}. These nuclear matrix elements arise from subleading terms in the relativistic expansion of the weak nuclear current, typically suppressed by powers of $1/M$ or $q/M$, where $M$ is the nucleon mass and $q$ is the momentum transfer. These operators, such as $p/M$ or $r.\sigma$, are relativistic in nature and generally lead to small matrix elements due to this suppression. In contrast, other operators such as the radial dipole term $rY_{lm}$ or the spin-orbital term $\sigma \times r$ emerge from leading-order non-relativistic contributions. These typically yield larger matrix elements, especially when they involve coherent contributions from many nucleons \cite{behrens1982, Dag_thesis}. The CVC-constrained s-NME values, derived from the above relation using the calculated l-NME values, are shown in Table \ref{nmes}.
The s-NME for the second forbidden decay of $^{93}$Zr and $^{135}$Cs is $^V\mathcal{M}^{(0)}_{211}$, for the  third forbidden decay of $^{85}$Br and $^{87}$Rb is  $^V\mathcal{M}^{(0)}_{321}$, for the fourth forbidden decay of $^{97}$Zr, $^{101}$Mo, $^{115}$Cd, $^{117}$Cd and $^{119}$In it is $^V\mathcal{M}^{(0)}_{431}$ and  for the fifth forbidden decay of $^{123}$Sn it is  $^V\mathcal{M}^{(0)}_{541}$. The CVC relation is very important and has been utilized in various studies, resulting in a significant enhancement of the  properties of beta decay. 
The magnitude of leading order s-NME $^V\mathcal{M}^{(0)}_{KK-11}$ is smaller than the leading order large vector l-NME $^V\mathcal{M}^{(0)}_{KK0}$. However, due to the systematic, order-by-order expansion introduced by Behrens and B$\ddot{\rm u}$hring \cite{behrens1982}, their contributions to $M_{K}(k_{e},k_\nu)$ and 
$m_{K}(k_{e},k_\nu)$ matrix elements in Eq. \ref{eq3} are on similar ground. This is because of the smaller phase space factors multiplying the $^V\mathcal{M}^{(0)}_{KK0}$ matrix element \cite{behrens1982}. 
%{\color{red}In Donnelly and Walecka formalism this can be explained without the help of phase space factor, see Ref. \cite{Ayala2022}.}

We further employ this s-NME to determine the electron spectral-shape as a function of $g_{A}$ for the investigated transitions.

\subsection{Sensitivity of electron spectral-shapes on $g_{A}$ values} 

We have computed the electron spectral-shapes for $^{85}$Br, $^{87}$Rb, $^{93}$Zr, $^{97}$Zr, $^{101}$Mo, $^{115}$Cd, $^{117}$Cd, $^{119}$In, $^{123}$Sn and $^{135}$Cs nuclei as a function of axial-vector coupling constant $g_{A}$.
For the computation of spectral-shapes, we have performed two sets of calculations. In the first set of calculation, the value of s-NME is zero or directly taken from SM, and in second one, s-NME is determined through the CVC relation. In the Figs. \ref{Fig1}-\ref{Fig4}, the first set of calculations are labeled as ``interaction name", while the second CVC-constrained calculations are denoted as ``interaction name + CVC". The CVC hypothesis assumes an `ideal' calculation within an infinite single-particle model space, incorporating all two-body and higher-order correlations. The electron spectral-shapes correspond to the integrand of Eq. \ref{eq2} are depicted as a function of the electron kinetic energy and the area under each curve to unity to make it easy to compare the experimental data in future experiments. The normalized spectra depend only on the ratio $g_V/g_A$, so adjusting $g_A$ alone is sufficient to get a better picture of electron spectra.

The electron spectral-shapes of the studied forbidden non-unique transitions show a dependency on $g_{A}$ values as well as on s-NMEs. Based on the shape of electron spectra, we have divided the studied transitions into three categories. {\bf Category I:} when electron-spectral shape is weakly dependent on $g_{A}$ values. In Fig. \ref{Fig1},  for $^{93}$Zr$(5/2^+) \rightarrow $$^{93}$Nb($9/2^+$) transition, the electron spectral shape is weakly dependent on $g_{A}$ values after constraining the s-NMEs. The dependency can be seen only in the lower electron kinetic energy region. After the electron kinetic energy around  0.02 MeV, the electron spectra become almost insensitive to $g_{A}$ values. 
  {\bf Category II:} when the spectrum is like a bell-shaped. In Fig. \ref{Fig2}-\ref{Fig3}, electron-spectral shapes of the third forbidden non-unique transition of nuclei $^{85}$Br, and $^{87}$Rb  and fourth forbidden non-unique transitions of $^{97}$Zr, $^{115}$Cd, $^{117}$Cd and fifth forbidden non-unique transition of $^{123}$Sn are shown. For $^{85}$Br$(3/2^-) \rightarrow ^{85}$Kr($9/2^+$), we see a similar type of $g_{A}$ dependency  before and after constraining s-NME calculations. For $^{87}$Rb$(3/2^-) \rightarrow ^{87}$Sr($9/2^+$), the electron spectral-shape is $g_{A}$ sensitive and after constraining the s-NMEs the $g_{A}$ dependency is still there and the shapes are similar in both the calculations. Previously, this spectra is studied using SM and compared with the MQPM in Ref. \cite{Joel_2017_2} but in the present study we also discuss the results after constraining the s-NME. For $^{97}$Zr$(1/2^+) \rightarrow ^{97}$Nb($9/2^+$) is  $g_{A}$ dependent  before and after constraining the s-NME. For the $^{115}$Cd$(1/2^+) \rightarrow ^{115}$In($9/2^+$) case, we are not getting any tilt in the shape of the electron spectrum before constraining s-NMES and after that we get bell-shaped structure. For the $^{117}$Cd$(1/2^+) \rightarrow ^{117}$In($9/2^+$) we see again $g_{A}$ dependency before and after CVC constrained calculations. The fifth forbidden non-unique decay $^{123}$Sn$(11/2^-)\rightarrow ^{123}$Sb($1/2^+$) is experimentally not measurable but still important from a theoretical point of view. The SM electron spectra is $g_{A}$ sensitive before and after constraining the s-NMEs, while in MQPM model \cite{Joel_2017}, they do not see any significant $g_{A}$ dependency for this transition. 
{\bf Category III:} when shape of the spectrum is neither a bell-shaped nor $g_{A}$ independent after constraining the s-NME. In Fig. \ref{Fig4},  the second forbidden non-unique transition of $^{135}$Cs and fourth forbidden non-unique transitions of $^{101}$Mo and $^{119}$In nuclei are shown. %^{101}$Mo$(1/2^+) \rightarrow ^{101}$Tc($9/2^+$)  and  $^{119}$In$(9/2^+) \rightarrow ^{119}$Sn($1/2^+$). 
For $^{135}$Cs$(7/2^+) \rightarrow $$^{135}$Ba($3/2^+$), the shape is $g_{A}$ independent before constraining the s-NME and the similar pattern can be seen using MQPM model \cite{Joel_2017}, but in our study the spectra after constraining s-NME shows  different behavior. So, we see our result match with MQPM when s-NMEs are zero and diverge when CVC is applied. {\color{black} It is important to mention here that CVC is essential for a proper determination of s-NMEs in both approaches.}
%In MQPM  \cite{Joel_2017}, the valence space was used for neutrons of reference nuclei for  A$ > $90 with single-particle-orbitals $1f_{5/2}-1i_{13/2}$, but in the present SM calculations, our model space does not include the $1i_{13/2}$ orbital, and may be because of lack of this orbital our electron spectrum is showing different shape.
The fourth forbidden non-unique decay of $^{101}$Mo$(1/2^+) \rightarrow ^{101}$Tc($9/2^+$) decay is weakly dependent on $g_{A}$ values before constraining the s-NME, while, after constraining the s-NME, the spectral-shape shows completely different behavior. The shape shows two humps in the throughout energy range, first hump which is in low energy region, is scattered and the second one is in higher energy range. The electron spectral-shape of $^{119}$In$(9/2^+) \rightarrow ^{119}$Sn($1/2^+$) decay is insensitive to the $g_{A}$ values throughout the energy range of electron kinetic energy before constraining s-NME and when s-NME is constrained then spectrum shows two peaks, one in lower energy side and other one in higher energy side, in general, with the increase of $g_{A}$ values shape becomes flattened in medium energy range and get a sudden fall in higher energy region. 

 {\color{black} In our present study we have not included NLO radiative corrections as done for NMEs calculations. With the radiative corrections in the LO we have done calculations for four transitions, $^{85}$Br$(3/2^-)\rightarrow ^{85}$Kr($9/2^+$), $^{97}$Zr$(1/2^+)\rightarrow ^{97}$Nb($9/2^+$), $^{135}$Cs$(7/2^+)\rightarrow $$^{135}$Ba($3/2^+$) and  $^{119}$In$(9/2^+) \rightarrow $$^{119}$Sn($1/2^+$) with $g_V=1.0$ and $g_A=1.0$ and 1.27. We found a very minimal change in the electron-spectrum figures. So, the radiative corrections do not significantly affect the final results and the shape of the spectra for the cases of studied transitions.}

\subsection{log$ft$ values} 
Table \ref{logft} presents the calculated log$ft$ values using both ``SM" and ``SM + CVC" approaches. The experimental data is available only in the cases of $^{93}$Zr, $^{135}$Cs, and $^{87}$Rb.
{\color{black} Constraining s-NMEs improves agreement for $^{93}$Zr and $^{135}$Cs, but not for $^{87}$Rb.}
In the case of the second forbidden non-unique decay of $^{93}$Zr, the branching ratio (BR) is 27$\%$. After constraining the s-NME, the moderate quenched log$ft$ value 11.847 is closer to the experimental value 12.10(10). For the second forbidden non-unique decay of $^{135}$Cs, the branching ratio is 100$\%$ and the experimental log$ft$ value is 13.48(6), the SM results are improved after constraining the s-NME. In the case of the third forbidden non-unique decay of $^{87}$Rb, having a branching ratio 100$\%$, the SM results with CVC are far from the experimental value 17.514(7). For the rest of the transitions, the experimental data is not available to compare the theoretical SM results. The other model MQPM and IBFM-2 are also needed to compare our SM  log$ft$ results. Our predicted  log$ft$ values are quite useful for future experiments.

\subsection{Decomposition of the integrated shape factor} 
For a comprehensive analysis of the integrated shape function, we have also calculated the values of $\tilde{C}$ along with its vector $\tilde{C}_V$, axial-vector $\tilde{C}_A$, and mixed vector-axial vector  $\tilde{C}_{VA}$, components for the studied transitions, as shown in Table \ref{decomposition}. The evaluation of these components follows from Eq. \ref{intc}.
We present all composition components with s-NMEs derived from CVC theory. Results where s-NMEs have zero value are labeled as ``SM”, while those incorporating s-NMEs from CVC theory are labeled as ``SM+CVC".
These results are given for  $g_{A}$=$g_{V}$=1.0. In all calculations, we have noticed that  the  sign of  $\tilde{C}_V$  and  $\tilde{C}_A$ is positive in both sets of the calculations. The sign of 
$\tilde{C}_{VA}$ component in ``SM" and ``SM+CVC" calculations varies. 
%{\color{red} The positive sign of  $\tilde{C}_A$ and $\tilde{C}_V$ arise from squared terms and $\tilde{C}_{VA}$ results from their interference, as per their definitions, see Ref. \cite{Ayala2022}.} 
For second forbidden non-unique $^{93}$Zr, the contribution of  axial-vector and vector components $\tilde{C}_A$ and $\tilde{C}_V$ before constraining s-NMEs is nearly similar,  while after constraining the s-NME $\tilde{C}_V$ dominates than $\tilde{C}_A$ and the sign of mixed component is opposite.
For $^{135}$Cs, the axial-vector and vector components $\tilde{C}_A$ and $\tilde{C}_V$ contribute nearly equally in both calculations. For $^{85}$Br, $^{87}$Rb, $^{97}$Zr, $^{101}$Mo,$^{115}$Cd and $^{119}$In, the SM calculations give dominant contribution of $\tilde{C}_V$ in comparison to $\tilde{C}_A$ while ``SM+CVC" calculations gives variation. For $^{101}$Mo and $^{119}$In the $\tilde{C}_V$ component dominant in comparison to $\tilde{C}_A$,  while other transitions show reciprocal behavior in ``SM+CVC" calculations.
For $^{85}$Br, $^{87}$Rb, $^{97}$Zr, $^{115}$Cd and $^{123}$Sn, the ``SM+CVC" shows the positive sign of 
mixed vector-axial vector $\tilde{C}_{VA}$ also they have a similar type of beta spectrum shape in the calculations after constraining the s-NME.  The $\tilde{C}_A$ term does not change its value when the s-NME is included because the s-NME is a vector nuclear matrix element so it affect only vector components and mixed component. The 3rd forbidden non-unique $^{87}$Rb$(3/2^-) \rightarrow 
 ^{87}$Sr($9/2^+$) transition is studied earlier using SM in Ref. \cite{Joel_2017_2} but in this paper the study of s-NME is not included in which the sign of $\tilde{C}_{VA}$ component is negative but in present study ``SM+CVC" gives positive sign of $\tilde{C}_{VA}$ component although the beta spectrum shows the similar pattern with respect to different $g_{A}$ values.

 %It is important to mention here, the radiative corrections at higher order do not significantly affect the shape of the spectra and values of decomposition of shape factor for the cases of studied transitions. 

 It is important to mention here that in the Donnelly and Walecka framework, the s-NME corresponds to the longitudinal multipole (LV), while the l-NME corresponds
to the Coulomb multipole (CV). Since these NMEs are of similar magnitude \cite{Ayala2022}, their similar contribution is straightforward in this formalism without requiring
the phase space factors.

\section{Conclusion} \label{Conclusion}
In the present manuscript, a systematic SM study of electron spectral-shapes, shape factors, and log$ft$ values has been performed for higher forbidden non-unique $\beta^{-}$ transitions of $^{85}$Br, $^{87}$Rb, $^{93}$Zr, $^{97}$Zr, $^{101}$Mo, $^{115}$Cd, $^{117}$Cd, $^{119}$In, $^{123}$Sn and $^{135}$Cs nuclei in the different mass region of nuclear chart. The nuclear matrix elements are calculated using well-known effective interactions in the SM Hamiltonian to predict the theoretical electron spectrum and other properties. The present manuscript mainly focuses on the sensitivity of electron spectral-shapes with different values of the axial-vector coupling constant.
We have found that electron spectral-shapes of second forbidden non-unique transition of $^{135}$Cs, third forbidden non-unique transition of $^{85}$Br and  $^{87}$Rb, fourth forbidden non-unique transition of $^{97}$Zr, $^{101}$Mo, $^{115}$Cd, $^{117}$Cd, $^{119}$In and fifth forbidden non-unique transition of $^{123}$Sn are strongly sensitive on $g_{A}$ values after constraining the s-NME from CVC theory. The second forbidden non-unique transition of $^{93}$Zr is weakly sensitive on $g_{A}$ after constraining s-NME from CVC theory. The transitions of $^{135}$Cs, $^{101}$Mo and $^{119}$In have been affected significantly with the s-NME and $g_{A}$ values. 
The second forbidden non-unique transition $^{135}$Cs$(7/2^+) \rightarrow $$^{135}$Ba($3/2^+$) has a branching ratio 100\% and has been experimentally observed, so this is an excellent candidate for SSM. This spectrum agrees well with the previously calculated spectrum with the MQPM  model \cite{Joel_2017} when s-NME is zero and diverges when constrained from CVC theory. For this transition,  in CVC calculation, spectrum get similar shape when $g_{A}$=1.0 and $g_{V}$=1.0 and at neighbouring values, the spectrum changes its shape.  
From experimental point of view, the third forbidden non-unique transition $^{87}$Rb$(3/2^-)\rightarrow $$^{87}$Sr($9/2^+$) is also important which has a branching ratio 100\% and it is also a good candidate for SSM. Previously, the electron spectrum corresponding to this transition has been calculated with   SM and MQPM models, but in the present study, we have also explored the role of s-NME \cite{Joel_2017}. The fourth forbidden non-unique transition of $^{101}$Mo and $^{119}$In is very different from the rest of the fourth forbidden non-unique transitions, the structure changes from bell-shaped to double hump like structure and we can see a significant role of s-NME in the spectrum. 
 
 Finally, in the present manuscript, we have constrained the s-NME directly using CVC theory and found that the computed electron spectral- shapes and shape factor values are sensitive to the value of s-NME. To see a better picture of electron spectral -shapes we have decomposed the total integrated shape function into its components. The signs and relative magnitudes of $C_{A}$, $C_{V}$, and $C_{VA}$ terms have been analyzed. The electron spectral-shapes exhibit sensitivity to the $g_{A}$ values and s-NME. Our theoretical study of electron spectral-shapes and exploration of s-NME with different $g_{A}$ values might be useful for future measurements of electron spectral-shapes in the context of SSM.

 \section*{Acknowledgement}
 %\vspace{-0.5cm}
	This work is supported by a research grant from SERB (India), CRG/2022/005167.\\
	We also acknowledge the National Supercomputing Mission (NSM) for providing computing resources of ‘PARAM Ganga’ at the Indian Institute of Technology Roorkee. We would also like to thank Prof. Jouni Suhonen and Dr. Anil Kumar for the useful discussions.

	%\newpage


\begin{thebibliography}{44}
		\expandafter\ifx\csname natexlab\endcsname\relax\def\natexlab#1{#1}\fi
		\expandafter\ifx\csname bibnamefont\endcsname\relax
		\def\bibnamefont#1{#1}\fi
		\expandafter\ifx\csname bibfnamefont\endcsname\relax
		\def\bibfnamefont#1{#1}\fi
		\expandafter\ifx\csname citenamefont\endcsname\relax
		\def\citenamefont#1{#1}\fi
		\expandafter\ifx\csname url\endcsname\relax
		\def\url#1{\texttt{#1}}\fi
		\expandafter\ifx\csname urlprefix\endcsname\relax\def\urlprefix{URL }\fi
		\providecommand{\bibinfo}[2]{#2}
		\providecommand{\eprint}[2][]{\url{#2}}

\bibitem{EJIRI20191}
H. Ejiri, J. Suhonen and K. Zuber,
``Neutrino–nuclear responses for astro-neutrinos, single beta decays and double beta decays",
\href{https://www.sciencedirect.com/science/article/pii/S0370157318303570}
{Phys. Rep. {\bf 797}, 1 (2019).}
        
\bibitem{Suhonen1} J. T. Suhonen,``Value of the Axial-Vector Coupling Strength in $\beta$ and $\beta\beta$ Decays: A Review ",
\href{https://doi.org/10.3389/fphy.2017.00055}
{Front. Phys. {\bf 5}, 55 (2017).}
    
\bibitem{Joel_2017}
  J. Kostensalo, M. Haaranen, and J. Suhonen, 
  ``Electron spectra in forbidden $\ensuremath{\beta}$ decays and the quenching of the weak axial-vector coupling constant ${g}_{A}$",
\href{https://link.aps.org/doi/10.1103/PhysRevC.95.044313}
      {Phys. Rev. C {\bf 95}, 044313 (2017).}

\bibitem{Joel_2017_2}
  J. Kostensalo, and  J. Suhonen, 
 ``${g}_{\mathrm{A}}$-driven shapes of electron spectra of forbidden $\ensuremath{\beta}$ decays in the nuclear shell model",
 \href {https://link.aps.org/doi/10.1103/PhysRevC.96.024317}
{Phys. Rev. C {\bf 96}, 024317 (2017).}

\bibitem{ED2007}
E. D. Commins, $Weak Interactions$ (McGraw-Hill, New York, 2007).

 \bibitem{jouni}
J. Suhonen, {\it From Nucleons to Nucleus: Concept of Microscopic Nuclear Theory},
 (Springer, Berlin 2007).
 
 \bibitem{Gysbers}
P. Gysbers et. al.,``Discrepancy between experimental and theoretical $\beta$-decay rates resolved from first principles",
\href{https://doi.org/10.1038/s41567-019-0450-7}
{Nature Physics 15, 428–431 (2019).}
 

 \bibitem{Martinez1996}
G. Mart\'{\i}nez-Pinedo,  A. Poves,  E. Caurier,  A. P. and Zuker, ``Effective ${g}_{A}$ in the $\mathrm{pf}$ shell",
\href{https://link.aps.org/doi/10.1103/PhysRevC.53.R2602}
{Phys Rev  C. {\bf 53}, R2602 (1996).}

\bibitem{Akumar1}
A. Kumar, P.C. Srivastava, and T. Suzuki, 
``Shell model results for nuclear $\beta^{-}$-decay properties of $sd$-shell nuclei",
\href{https://doi.org/10.1093/ptep/ptaa012}
{Prog. of Theor.  Exp. Phys. {\bf 2020}, 033D01 (2020).}

\bibitem{Vkumar1}
V. Kumar, P.C. Srivastava, and H. Li, 
``Nuclear $\beta^{-}$-decay half-lives for $fp$ and $fpg$ shell nuclei,
\href{https://ui.adsabs.harvard.edu/abs/2016JPhG...43j5104K}
{J. Phys. G:Nucl. Part. Phys. {\bf 43}, 105104 (2016).}

 \bibitem{Ejiri2014}
H. Ejiri, N. Soukouti and J. Suhonen,``Spin-dipole nuclear matrix elements for double beta decays and astro-neutrinos",
\href{https://doi.org/10.1016/j.physletb.2013.12.051}
{Phys. Lett. B  {\bf 729}, 27 (2014).}

\bibitem{Ejiri2015}
H. Ejiri and J. Suhonen,``GT neutrino–nuclear responses for double
beta decays and astro neutrinos",
\href{ http://dx.doi.org/10.1088/0954-3899/42/5/055201}
{J. Phys. G: Nucl. Part. Phys. {\bf 42}, 055201 (2015).}



 \bibitem{mika2016}
M. Haaranen, P. C. Srivastava, and J. Suhonen
``Forbidden non-unique $\ensuremath{\beta}$ decays and effective values of weak coupling constants'',
\href{https://link.aps.org/doi/10.1103/PhysRevC.93.034308}
{Phys. Rev. C {\bf {93}}, 034308 (2016).}

\bibitem{Bodenstein2020}
L. Bodenstein-Dresler et al.(COBRA Collaboration),
``Quenching of $g_{A}$ deduced from the $\beta$-spectrum shape of $^{113}$Cd measured with the COBRA experiment", 
\href{https://doi.org/10.1016/j.physletb.2019.135092}
{Phys. Lett. B {\bf 800}, 135092(2020).}

\bibitem{joel2021}
J. Kostensalo, J. Suhonen, J. Volkmer, S. Zatschler, K. Zuber,
``Confirmation of $g_{A}$ quenching using the revised spectrum-shape method for the analysis of the $^{113}$Cd $\beta$-decay as measured with the COBRA demonstrator",
\href{https://doi.org/10.1016/j.physletb.2021.136652}
{Phys. Lett. B {\bf 822}, 136652(2021).}


\bibitem{Leder2022} A. F. Leder et al.,``Determining ${g}_{A}/{g}_{V}$ with High-Resolution Spectral Measurements Using a ${\mathrm{LiInSe}}_{2}$ Bolometer",
\href{10.1103/PhysRevLett.129.232502}
{Phys. Rev. Lett. {\bf 129}, 232502 (2022).}


\bibitem{Pagnanini 2023}
L. Pagnanini et al. (ACCESS Collaboration),``Array of cryogenic calorimeters to evaluate the spectral shape of forbidden $\beta$ -decays: the ACCESS project", 
\href{https://doi.org/10.1140/epjp/s13360-023-03946-x}
{Eur. Phys. J. A {\bf 138}, 445 (2023).}

\bibitem{Iachello}
F. Iachello and A. Arima, \textit{The Interacting Boson Model} (Cambridge University Press, Cambridge, 1987). 

\bibitem{Kotila}
M. Haaranen,  J. Kotila,  J. and Suhonen,
``Spectrum-shape method and the next-to-leading-order terms of the $\ensuremath{\beta}$-decay shape factor",
\href{https://link.aps.org/doi/10.1103/PhysRevC.95.024327}
{Phys. Rev. C {\bf 95}, 024327 (2017).}

\bibitem{Belli2007}
P. Belli et al.,``Investigation of \ensuremath{\beta} decay of $^{113}\mathrm{Cd}$",
\href{https://link.aps.org/doi/10.1103/PhysRevC.76.064603}
{Phys. Rev. C {\bf 76}, 064603 (2007).}



\bibitem{Anil2020PRC}
A. Kumar, P. C. Srivastava, J. Kostensalo and J. Suhonen,
``Second-forbidden nonunique ${\ensuremath{\beta}}^{\ensuremath{-}}$ decays of $^{24}\mathrm{Na}$ and $^{36}\mathrm{Cl}$ assessed by the nuclear shell-model'',
\href{https://link.aps.org/doi/10.1103/PhysRevC.101.064304}
{Phys. Rev. C. {\bf 101}, 064304 (2020).}
  
  \bibitem{anil2021}
A. Kumar,  P. C. Srivastava, and J. Suhonen,``Second-forbidden nonunique $\beta^{-}$ decays of $^{59,60}$Fe: possible
candidates for $g_{A}$ sensitive electron spectral-shape measurements",
\href{https://doi.org/10.1140/epja/s10050-021-00540-6}
{Eur. Phys. J. A  {\bf 57}, 225 (2021).}



  \bibitem{Gregorio2024} 
 G. De Gregorio,  R. Mancino,  L. Coraggio, and  N. Itaco,``Forbidden $\ensuremath{\beta}$ decays within the realistic shell model",
 \href{https://link.aps.org/doi/10.1103/PhysRevC.110.014324}
 {Phys. Rev. C {\bf 110}, 014324 (2024).}

\bibitem{Coraggio2019} 
 L. Coraggio,  L.  De Angelis, T. Fukui,  A.  Gargano, N. Itaco,  and  F. Nowacki,``Renormalization of the Gamow-Teller operator within the realistic shell model",
\href{https://link.aps.org/doi/10.1103/PhysRevC.100.014316}
{Phys. Rev. C {\bf 100}, 014316 (2019).}

\bibitem{Coraggio2022}
L. Coraggio, N. Itaco, G. De Gregorio, A. Gargano, R. Mancino, and F. Nowacki,``Shell-model calculation of $^{100}\mathrm{Mo}$ double-$\ensuremath{\beta}$ decay",
\href{https://link.aps.org/doi/10.1103/PhysRevC.105.034312}
{Phys. Rev. C {\bf 105}, 034312 (2022).}

  \bibitem{Ramalho2024}M. Ramalho, J. Suhonen, 
  ``${g}_{A}$-sensitive $\ensuremath{\beta}$ spectral shapes in the mass $A=86-99$ region assessed by the nuclear shell model",
  \href{https://link.aps.org/doi/10.1103/PhysRevC.109.034321}
 {Phys. Rev. C {\bf 109}, 034321 (2024).}
 
 \bibitem{Ramalho}
 M. Ramalho, J. Suhonen, 
``Computed total $\ensuremath{\beta}$-electron spectra for decays of Pb and Bi in the $^{220,222}\mathrm{Rn}$ radioactive chains",
 \href{https://link.aps.org/doi/10.1103/PhysRevC.109.014326}
 {Phys. Rev. C {\bf 109}, 014326 (2024).}

\bibitem{joel2023}
J. Kostensalo, E. Lisi, A. Arrone, and J. Suhonen,
``$^{113}\mathrm{Cd}$ $\ensuremath{\beta}$-decay spectrum and ${g}_{A}$ quenching using spectral moments",
\href{https://link.aps.org/doi/10.1103/PhysRevC.107.05550}
 {Phys. Rev. C {\bf 107}, 055502 (2023).}



 
\bibitem{mst2006}
M. T. Mustonen, M. Aunola, and J. Suhonen,
``Theoretical description of the fourth-forbidden non-unique $\beta$ decays of $^{113}$Cd and $^{115}$In'',
\href{https://link.aps.org/doi/10.1103/PhysRevC.73.054301}
{Phys. Rev. C {\bf {73}}, 054301 (2006)};
Erratum \href{https://link.aps.org/doi/10.1103/PhysRevC.76.019901}
{ Phys. Rev. C {\bf {76}}, 019901 (2007).}

  
  \bibitem{behrens1982}
H.~Behrens and W.~B$\ddot{{\rm u}}$hring,  {\it Electron Radial Wave Functions and Nuclear Beta-Decay} (Clarendon Press, 1982).


\bibitem{hfs1966}
H. F. Schopper, {\it Weak Interaction and Nuclear Beta Decay} (North-Holland, Amsterdam, 1966).



\bibitem{Holstein1974}
B. R. Holstein, 
``Recoil effects in allowed beta decay: The elementary particle approach",
\href{https://link.aps.org/doi/10.1103/RevModPhys.46.789}
{Rev. Mod. Phys. {\bf 46}, 789-814 (1974).}



\bibitem{Connell1972}
J. S. O'Connell, T. W. Donnelly,  and J. D. Walecka, 
``Semileptonic Weak Interactions with ${\mathrm{C}}^{12}$"
\href{https://link.aps.org/doi/10.1103/PhysRevC.6.719}
{Phys. Rev.  C. {\bf 6}, 719 (1972).}

\bibitem{DONNELLY1972275}
T. W. Donnelly and J. D. Walecka,
``Semi-leptonic weak and electromagnetic interactions in nuclei with application to $^{16}$O",
\href{https://www.sciencedirect.com/science/article/pii/0370269372905771}
{Phys. Lett. {\bf 41B}, 275 (1972).}

\bibitem{DONNELLY197381}
T. W. Donnelly and J. D. Walecka,
``Elastic magnetic electron scattering and nuclear moments",
\href{https://www.sciencedirect.com/science/article/pii/0375947473906891}
{Nucl. Phys. A {\bf 201}, 81 (1973).}


\bibitem{DONNELLY1976368}
T. W. Donnelly and J. D. Walecka,
``Semi-leptonic weak and electromagnetic interactions with nuclei: Isoelastic processes",
\href{https://www.sciencedirect.com/science/article/pii/0375947476902098}
{Nucl. Phys. A {\bf 274}, 368 (1976).}

\bibitem{DONNELLY19791}
T. W. Donnelly and R. D. Peccei,
``Neutral current effects in nuclei",
\href{https://www.sciencedirect.com/science/article/pii/0370157379900103}
{Phys. Rep. {\bf 50}, 1 (1979).}

\bibitem{Walecka}
J. D. Walecka,
{\it ``Theoretical Nuclear and Subnuclear Physics} (World Scientific, 2004)".

\bibitem{Ayala2022}
A. Glick-Magid and D. Gazit,
``A formalism to assess the accuracy of nuclear-structure weak interaction effects in precision $\beta$ -decay studies",
\href{https://iopscience.iop.org/article/10.1088/1361-6471/ac7edc}
{J. Phys. G: Nucl. Part. Phys. {\bf 49} 105105 (2022).}


\bibitem{Patrignani}
C. Patrignani and Particle Data Group,
``Review of Particle Physics'',
\href{https://doi.org/10.1088/1674-1137/40/10/100001}
{Chinese Phys. C {\bf 40}, 100001 (2016).}

    
\bibitem{KSHELL}
 N. Shimizu, T. Mizusaki, Y. Utsuno and Y. Tsunoda, 
``Thick-restart block Lanczos method for large-scale shell-model calculations",
\href{https://doi.org/10.1016/j.cpc.2019.06.011}
{Comput. Phys. Commun. {\bf  244}, 372–384 (2019).}


    
    \bibitem{gwbxg1}
    Z. Ren et al.,
    ``Reinvestigation of the level structures of the $N=49$ isotones $^{89}\mathrm{Zr}$ and $^{91}\mathrm{Mo}$",
    \href{https://link.aps.org/doi/10.1103/PhysRevC.106.024323}
    {Phys. Rev. C {\bf 106}, 024323 (2022).}

    \bibitem{gwbxg2}
    P. Dey et al.,
    ``Experimental investigation of high-spin states in $^{90}\mathrm{Zr}$",
    \href{https://link.aps.org/doi/10.1103/PhysRevC.105.044307}
    {Phys. Rev. C {\bf 105}, 044307 (2022).}
    
    \bibitem{G-matrix}
    N. Boelaert, N. Smirnova,  K. Heyde, and  J. Jolie,
    ``Shell model description of the low-lying states of the neutron deficient Cd isotopes"
    \href{https://link.aps.org/doi/10.1103/PhysRevC.75.014316}
     {Phys RevC {\bf 75}, 014316 (2007).}

        
    \bibitem{snet1}
    A. Hosaka, et al., 
    ``G-matrix effective interaction with the paris potential", 
   \href{https://www.sciencedirect.com/science/article/pii/0375947485902921}
   {Nucl. Phys. A {\bf 71}, 444 (1986).}
 
    \bibitem{snet2}
    B. A. Brown, et al., 
    ``Gamow-Teller strength in the region of $^{100}\mathrm{Sn}$",
    \href{https://link.aps.org/doi/10.1103/PhysRevC.50.R2270}
    {Phys. Rev. C {\bf 50}, 2270(R) (1994).}
   
    \bibitem{sn100pn1}
    B. A. Brown, N. J. Stone,  J. R. Stone,  I. S.  Towner, and M. Hjorth-Jensen,
    ``Magnetic moments of the ${2}_{1}^{+}$ states around $^{132}\mathrm{Sn}$",
    \href{https://link.aps.org/doi/10.1103/PhysRevC.71.044317}
    {Phys. Rev. C {\bf 71}, 044317 (2005).}
     
     \bibitem{sn100pn2}
     R. Machleidt, F. Sammarruca, and Y. Song,
     ``Nonlocal nature of the nuclear force and its impact on nuclear structure",
    \href{https://link.aps.org/doi/10.1103/PhysRevC.53.R1483}
     {Phys. Rev. C {\bf 71}, 53 R1483 (1996).}
 

 \bibitem{H7B}
  A. Hosaka, K.-I. Kubo and H. Toki,
  ``G-matrix effective interaction with the paris potential",
  \href{https://www.sciencedirect.com/science/article/pii/0375947485902921}
  {Nucl. Phys. A {\bf 444}, 76 (1985).}

\bibitem{Ji1988}
X. Ji and B. H. Wildenthal,
``Effective interaction for N=50 isotones",
\href{https://doi.org/10.1103/PhysRevC.37.1256}
{Phys. Rev. C {\bf 37}, 1256 (1988).}

\bibitem{Gloeckner1975}
D. H. Gloeckner,
``Shell-model systematics of the zirconium and niobium isotopes",
\href{https://doi.org/10.1016/0375-9474(75)90484-4}
{Nucl. Phys. A {\bf 253}, 301 (1975).}

\bibitem{Serduke1976}
 F. J. D. Serduke, R. D. Lawson, and D. H. Gloeckner,
 ``Shell-model study of the $N$ = 49 isotones",
 \href{https://doi.org/10.1016/0375-9474(76)90094-4}
 {Nucl. Phys. A {\bf 256}, 45 (1976).}           

\bibitem{Machleidt2001}
R. Machleidt,``High-precision, charge-dependent Bonn nucleon-nucleon potential",
\href{https://link.aps.org/doi/10.1103/PhysRevC.63.024001}
{Phys. Rev.  C. {\bf 63}, 024001 (2001).}

\bibitem{Behrens1971}
H. Behrens and W.~B$\ddot{{\rm u}}$hring, ``Nuclear beta decay",
\href{https://doi.org/10.1016/0375-9474(71)90489-1}
{Nucl. Phys. A {\bf 162}, 111 (1971).}

\bibitem{Sadler}
R. Sadler and H. Behrens,`` Second-forbidden beta-decay and the effect of (V+A)- and S-interaction admixtures:$^{36}$Cl",
\href{https://doi.org/10.1007/BF01290778}
{Z. Physik A {\bf 346}, 25 (1993).}
 
\bibitem{Zhi2013}
Q. Zhi, E. Caurier, J. J. Cuenca-Garc\'{\i}a, K. Langanke,  G. Mart\'{\i}nez-Pinedo, K. and Sieja,``Shell-model half-lives including first-forbidden contributions for $r$-process waiting-point nuclei",
\href{https://link.aps.org/doi/10.1103/PhysRevC.87.025803}
{Phys. Rev. C {\bf 87}, 025803 (2013).}

\bibitem{Dag_thesis}
%Dag Isak August Fahlin Str$\ddot{{\rm o}}$mberg,``Weak interactions in degenerate oxygen-neon cores", 
%\href{https://tuprints.ulb.tu-darmstadt.de/13302}
%{Ph.D. thesis (2020), GSI.}
Dag Isak August Fahlin Str$\ddot{{\rm o}}$mberg,``Weak interactions in degenerate oxygen-neon cores",  Ph.D. thesis (2020),
\href{https://tuprints.ulb.tu.darmstadt.de/13302}
{https://tuprints.ulb.tu.darmstadt.de/13302}.


\bibitem{AME2020}
M. Wang, W. J. Huang, F. G. Kondev, G. Audi and S. Naimi,
``The AME 2020 atomic mass evaluation (II). Tables, graphs and references"
\href{https://dx.doi.org/10.1088/1674-1137/abddaf}
{Chin. Phys. C {\bf 45}(3), 030003 (2021).}




 \end{thebibliography}
\end{document}